\RequirePackage[2020-02-02]{latexrelease}
\documentclass[aps,prd,superscriptaddress,showpacs,preprint]{revtex4}
\usepackage{graphicx, bm}
\usepackage[usenames]{color}
\usepackage{float}
\usepackage{multirow}
\usepackage{subcaption}

\usepackage{amsmath}

\usepackage[colorlinks=true,
            linkcolor=blue,
            citecolor=blue,
            urlcolor=blue]{hyperref}
\usepackage[capitalize,nameinlink]{cleveref}

\begin{document}

\draft
\title{Production of the top partner $T$ of the Bestest Little Higgs Model at the future muon collider}

\author{E. Cruz-Albaro\footnote{elicruzalbaro88@gmail.com}}
\affiliation{\small Facultad de F\'{\i}sica, Universidad Aut\'onoma de Zacatecas\\
            Apartado Postal C-580, 98060 Zacatecas, M\'exico.\\}

\author{D. Espinosa-G\'omez \footnote{david.espinosa@umich.mx}
}
\affiliation{\small Facultad de Ciencias F\'{\i}sico Matem\'aticas, Universidad Michoacana de San Nicol\'as de Hidalgo\\
            Avenida Francisco, J. M\'ujica S/N, 58060, Morelia, Michoac\'an, M\'exico.\\}

\author{A. Guti\'errez-Rodr\'{\i}guez\footnote{alexgu@fisica.uaz.edu.mx}}
\affiliation{\small Facultad de F\'{\i}sica, Universidad Aut\'onoma de Zacatecas\\
      Apartado Postal C-580, 98060 Zacatecas, M\'exico.\\}

\author{F. Ramírez-Zavaleta \footnote{iguazu.ramirez@umich.mx}
}
\affiliation{\small Facultad de Ciencias F\'{\i}sico Matem\'aticas, Universidad Michoacana de San Nicol\'as de Hidalgo\\
            Avenida Francisco, J. M\'ujica S/N, 58060, Morelia, Michoac\'an, M\'exico.\\}

\author{E. S. Tututi \footnote{eduardo.tututi@umich.mx}
}
\affiliation{\small Facultad de Ciencias F\'{\i}sico Matem\'aticas, Universidad Michoacana de San Nicol\'as de Hidalgo\\
            Avenida Francisco, J. M\'ujica S/N, 58060, Morelia, Michoac\'an, M\'exico.\\}

\author{A. M. Fernández-Figueroa \footnote{1638233b@umich.mx}
}
\affiliation{\small Facultad de Ciencias F\'{\i}sico Matem\'aticas, Universidad Michoacana de San Nicol\'as de Hidalgo\\
            Avenida Francisco, J. M\'ujica S/N, 58060, Morelia, Michoac\'an, M\'exico.\\}

\date{\today}

\begin{abstract}
In this work, we evaluate the decay widths and branching ratios of the heavy top partner $T$, and perform a comprehensive phenomenological study of its pair and associated production at a muon collider within the framework of the Bestest Little Higgs Model (BLHM).
 The BLHM provides an attractive and natural mechanism  for addressing the fine-tuning problem in the Standard Model (SM) of particle physics.
In our study of the  top partner $T$, we consider the production processes $\mu^{+} \mu^{-} \to (\gamma, Z, Z', h, H) \to T \bar{T}$ and $\mu^{+} \mu^{-} \to (\gamma, Z, Z', h, H)\to \bar{t}T+ t \bar{T}$, and we explore different center-of-mass energies of the muon collider to perform our numerical analysis of the production cross sections for the processes of interest. Our results show that pair production is the most promising channel for probing the top partner of the BLHM at a future muon collider. \\

\end{abstract}

\pacs{12.60.-i,  13.87.Ce, 14.65.Ha      \\
Keywords: Models beyond the standard model, Production, Top quarks. }

\vspace{5mm}

\maketitle


\section{Introduction}

Currently, the Large Hadron Collider (LHC) at CERN has delivered an unprecedented amount of data from proton–proton collisions, opening unique opportunities to address fundamental questions that remain unresolved in particle physics. These include, for example, the observed asymmetry between matter and antimatter, the nature of dark matter and dark energy, the  observations of neutrino oscillations, and theoretical issues such as the strong
CP problem~\cite{Dine:1986bg,Kim:2008hd}, the fine-tuning or naturalness problem~\cite{Susskind:1978ms}, among other things. Although the SM of particle physics is a remarkably successful theoretical framework, capable of accurately describing a wide range of experimental observations, it is nonetheless unable to account for several of these fundamental phenomena. Consequently, many of the outstanding puzzles of the SM strongly motivate the existence of new physics beyond the SM (BSM). Thus, the search for new physics remains a frontier in particle physics research.

To address the hierarchy and fine-tuning problems of the SM, several extensions have been put forward that introduce the existence of new particles, in particular heavy partners of the SM quarks, also known as vector-like quarks~\cite{Hill:2002ap,Perelstein:2003wd,Matsedonskyi:2012ym,Buchkremer:2013bha,Dimopoulos:1981xc,Arkani-Hamed:2001nha}. These new heavy quarks are hypothetical spin-1/2 particles that play a crucial role in reducing the quadratically divergent corrections to the Higgs boson mass, which arise at one or more loop levels from SM fermions and gauge bosons. Consequently, these new particles are expected to couple preferentially to the third-generation SM quarks, since in this context the top-quark loop provides the largest divergent contribution to the Higgs mass. Among the theories that predict the existence of vector-like quarks is the BLHM~\cite{Schmaltz:2010ac}, which gives rise to several new fermionic states, including four heavy top partners ($T, T_5, T_6, T^{2/3}$), a heavy bottom partner ($B$), and an exotic quark with electric charge $+5/3$ ($T^{5/3}$).

The BLHM is a relatively recent model that provides an elegant solution to the hierarchy problem through collective symmetry breaking, thereby eliminating the need for fine-tuning, as required in the SM. The BLHM was also formulated to address certain theoretical inconsistencies~\cite{Schmaltz:2010ac,Schmaltz:2008vd} present in most other formulations of little Higgs models (such as the simplest little Higgs~\cite{Schmaltz:2004de}, the little Higgs model and custodial $SU(2)$~\cite{Chang:2003un},  and the littlest Higgs models~\cite{Arkani-Hamed:2002ikv}, among others). 
For instance, conventional little Higgs models typically exhibit tension with precision electroweak observables in the gauge sector. In contrast, the BLHM introduces two distinct symmetry-breaking scales, $f$ and $F$, with $F>f$. These scales represent the energy scale of new physics, such that the new heavy quarks acquire masses proportional to the scale $f$, while the new gauge bosons obtain squared masses proportional to $f^{2}+F^{2}$, thereby reducing their contributions to precision electroweak observables.
Within the same framework, a satisfactory Higgs quartic coupling is generated without introducing any dangerous singlet fields in the scalar sector, as occurs in other little Higgs models. Instead, an additional singlet  is introduced, carrying specific electroweak quantum numbers and preserving a custodial $SU(2)$ symmetry.  For further complementary studies of the BLHM, see Refs.~\cite{Kalyniak:2013eva,Cruz-Albaro:2024vjk,Cervantes-Baltazar:2025cni}.

At the LHC, new heavy quarks such as the exotic charge $+5/3$ particles $T^{5/3}$, as well as heavy up- and down-type quarks, $T$ and $B$, are being extensively searched for over a wide range of energies~\cite{ATLAS:2026ojf,CMS:2024bni,ATLAS:2023pja,ATLAS:2024fdw,ATLAS:2022tla,CMS:2018zkf,CMS:2018ubm,ATLAS:2017vdo}. These particles can be produced predominantly in pairs through the strong interaction or singly via electroweak interactions~\cite{ATLAS:2024fdw}. The latter production mode becomes dominant for quark masses above 1 TeV due to the larger available phase space.
Recently, the  CMS~\cite{CMS:2024bni} and ATLAS~\cite{ATLAS:2024fdw,ATLAS:2022tla}  collaborations have reported results from searches for the production of these quarks. The analyses are based on data collected by the corresponding experiments during the Run 2 of the LHC in proton–proton collisions at a center-of-mass energy of 13 TeV, corresponding to an integrated luminosity of 138 fb$^{-1}$ recorded at CERN  between 2016 and 2018.
In the aforementioned experiments, searches for these new quarks have been performed in final states containing a Higgs boson or a gauge boson ($Z$, $W$) decaying hadronically in association with top and bottom quarks, as well as in multilepton events and events with a single lepton accompanied by a large number of jets, among other signatures. In general, the production and decay of heavy quarks have been investigated across a wide range of possible final states.

In this paper, we explore the phenomenology of pair production and single production (in association with the SM top quark) of the top partner $T$ through the processes   $\mu^{+} \mu^{-} \to (\gamma, Z, Z', h, H) \to T \bar{T}$ and  $\mu^{+} \mu^{-} \to (\gamma, Z, Z', h, H)\to \bar{t} T+ t \bar{T}$  within the framework of the BLHM  at a muon collider.
It is worth noting that the mediator particles $Z'$ and $H$ correspond to new gauge and Higgs bosons, respectively,  predicted by the extended model under consideration. In contrast, the light Higgs boson $h$ exhibits properties similar to those of the SM Higgs boson.
  Our study is carried out in the context of a muon collider, as this future collider provides an exceptionally clean background environment and could operate at high energies in the multi-TeV range~\cite{Accettura:2023ked,Long:2020wfp}. This would offer a significant opportunity to open an unprecedented energy threshold for new physics and could potentially enable highly precise measurements of processes that remain inaccessible at current colliders.
On the other hand, the search for a heavy partner of the top quark has also been explored in other extended scenarios such as the little Higgs models~\cite{Harigaya:2011yg}, the little Higgs model with T-parity~\cite{Kong:2007uu}, the supersymmetric theory with R-parity~\cite{Kong:2007uu},  the supersymmetric standard model~\cite{Kitano:2002ss}, the left-right twin Higgs model~\cite{Liu:2014pts}, among others. 

The article is organized as follows.  
In~\cref{sec:TT}, we determine the production cross sections for the pair and single production of the heavy top-quark partner.
In~\cref{sec:results}, the numerical results are discussed. Finally, conclusions are presented in~\cref{sec:conclusions}. 
\cref{app:amplitudes} provides the different transition amplitudes for the pair and single production of the heavy quark $T$ through the production channels $\mu^{+} \mu^{-} \to  T \bar{T} $ and $\mu^{+} \mu^{-} \to \bar{t} T + t \bar{T}$. 
\cref{app:rulesF} presents the effective couplings, as well as the vector and axial-vector couplings, involved in our calculations.

\section{ Production processes $\mu^{+} \mu^{-} \to  \bar{T} T $ and $\mu^{+} \mu^{-} \to \bar{t} T + t \bar{T}$} \label{sec:TT}

\begin{figure}[H]
    \center
 \subfloat[]{\includegraphics[width=7.10cm]{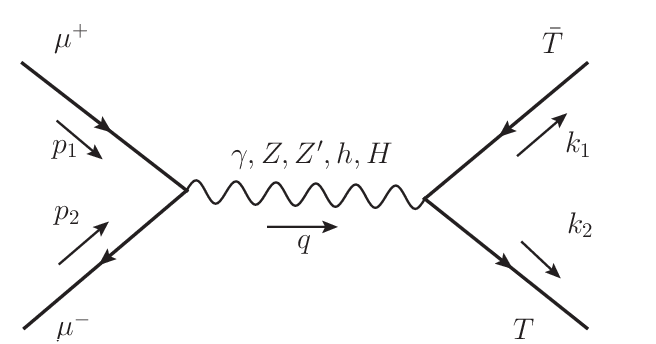}}
\subfloat[]{\includegraphics[width=7.10cm]{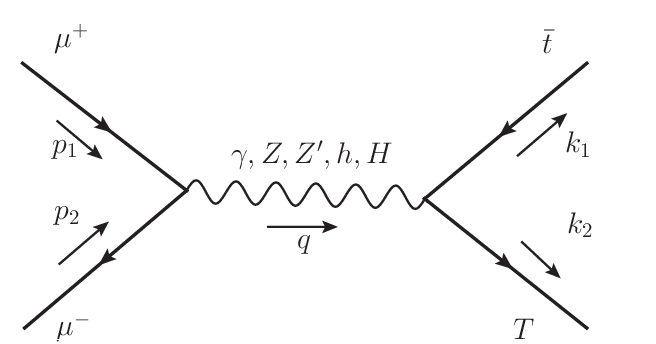}}\\
\subfloat[]{\includegraphics[width=7.10cm]{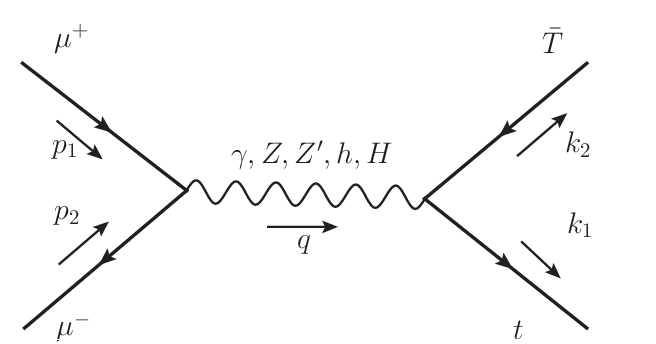}} 
    \caption{ Feynman diagrams for production processes  of the top partner $T$ at the muon collider: a)  $\mu^+\mu^-\rightarrow T\bar{T}$,  b) $\mu^+\mu^-\to  \bar{t}T$, and c) $\mu^+\mu^-\to t \bar{T} $.}
\label{fig:TT}
\end{figure}

In this section, we determine the production cross sections of the heavy top-quark partner. For this purpose, \cref{fig:TT} presents the Feynman diagrams for both pair production and single production of the top partner $T$ at a muon collider within the framework of the BLHM. From these Feynman diagrams, we derive the corresponding scattering amplitudes for each mediator particle (see \cref{eq:MTTi,eq:MTTi1,eq:MTTi2,eq:MTTi3,eq:MTTf}  in~\cref{app:amplitudes}), which are subsequently used to obtain $\sigma^{T\bar{T}}_{i}$,  $\sigma^{\bar{t} T}_{i}$, and $\sigma^{t \bar{T}}_{i}$ representing the production cross sections of the heavy quark $T$ when the mediator particles are the particles $i$, where $i\equiv \gamma, Z, Z', h, H, \gamma Z Z' h H$.
 Specifically, the cross sections $\sigma^{T\bar{T}}_{\gamma Z Z' h H}$, $\sigma^{\bar{t} T}_{\gamma Z Z' h H}$, and $\sigma^{ t \bar{T}}_{\gamma Z Z' h H}$ represent the interference terms between the bosons involved.

\subsection{ Process $\mu^+\mu^-\to T\bar{T}$ }

Regarding the pair production of the heavy quark $T$ through the  $\mu^+\mu^-\to (\gamma, Z, Z', h, H ) \to T\bar{T}$ process, its total cross section is determined according to \cref{tot-TT}:

\begin{align} \label{tot-TT}
    \sigma_{total}(\mu^+\mu^-\to T\bar{T})=\sigma_{\gamma}^{T\bar{T}}+\sigma_{Z}^{T\bar{T}}+\sigma_{Z'}^{T\bar{T}} +\sigma_{h}^{T\bar{T}} +\sigma_{H}^{T\bar{T}} +\sigma_{\gamma ZZ'h H}^{T\bar{T}},
\end{align}

\noindent where

\begin{align} \label{TT1}
    \sigma_{\gamma}^{T\bar{T}}=\frac{\left(g_{V}^{\mu\mu\gamma}\right)^2\lambda}{12\pi s^2}\left((g_A^{T\bar{T}\gamma})^2\lambda^2s+(g_V^{T\bar{T}\gamma})^2(6m_T^2+\lambda^2s)\right),
\end{align}

\begin{align} \label{width-z}
    \sigma_Z^{T\bar{T}}=\frac{(g_V^{T\bar{T}Z})^2\lambda(6m_T^2+\lambda^2s)+(g_A^{T\overline{T}Z})^2 \lambda^3 s}{12\pi\left(m_Z^4+m_Z^2(\Gamma_Z^2-2s)+s^2\right)}\left((g_A^{Z \mu\mu})^2+(g_V^{Z \mu\mu })^2\right),
\end{align}

\begin{align} \label{width-z1}
   	\sigma_{Z'}^{T\bar{T}}=\frac{(g_V^{T\bar{T}Z'})^2\lambda(6m_T^2+\lambda^2s)+(g_A^{T\overline{T}Z'})^2 \lambda^3 s}{12\pi\left(m_{Z'}^4+m_{Z'}^2(\Gamma_{Z'}^2-2s)+s^2\right)}\left((g_A^{Z' \mu\mu})^2+(g_V^{Z' \mu\mu })^2\right),
\end{align}

\begin{equation} \label{width-z2}
    \sigma_h^{T\bar T}=\frac{1}{16\pi((s-m_h^2)^2+m_h^2\Gamma_h^2)}\left(g_{hT\bar{T}}^2g_{h\mu\mu}^2s\lambda^3\right),
\end{equation}

\begin{equation} \label{width-z3}
    \sigma_H^{T\bar T}=\frac{1}{16\pi((s-m_H^2)^2+m_H^2\Gamma_H^2)}\left(g_{HT\bar{T}}^2g_{H\mu\mu}^2s\lambda^3\right),
\end{equation}


\begin{align} \label{TT2}
	\sigma_{\gamma ZZ'h H}^{T\bar{T}}=&\frac{\lambda}{24\pi}  \bigg [4 g_V^{\mu\mu\gamma}
		\bigg(\frac{g_{A}^{T\bar{T}\gamma} g_{V}^{Z'\mu\mu} g_A^{T\bar{T}Z'} \lambda^2
			\left(s-m_{Z'}^2\right)}{\Gamma_{Z'}^2 m_{Z'}^2+\left(m_{Z'}^2-s\right)^2}-\frac{g_V^{T\bar{T}\gamma} g_{V}^{Z\mu\mu} g_V^{T\bar{T} Z}
			\left(m_Z^2-s\right) \left(6 m_{T}^2+\lambda ^2 s\right)}{s\left(\Gamma_Z^2
			m_Z^2+\left(m_Z^2-s\right)^2\right)}\bigg)\nonumber\\
		&+\frac{3 g_{h\mu \mu} g_{H\mu\mu} g_{hT\bar{T}} g_{HT\bar{T}} \lambda^2
			s \left(\left(s-m_h^2\right)
			\left(s- m_H^2\right)+\Gamma_h \Gamma_H m_h m_H\right)}{\left(\left(s-m_h^2\right)^2  + \Gamma_h^2 m_h^2 \right) \left(\left(s-m_H^2\right)^2 + \Gamma_H^2 m_H^2 \right)}\bigg].
\end{align}

\noindent In these expressions,  $   g_{h T\bar{T} }, g_{h\mu\mu}, g_{H\mu\mu}$, and  $g_{HT\bar{T}}$ denote coupling constants; $\sqrt{s}$ represents the center-of-mass energy, and $g^{x_i}_{V}$ and $g^{x_i}_{A}$  (with $x_i \equiv T\bar{T}\gamma, T\bar{T}Z, T\bar{T}Z',  \mu\mu \gamma,  Z\mu \mu, Z' \mu \mu $) correspond to the vector and axial-vector couplings, respectively, which are explicitly provided in \cref{Table-i,Table-ii,Table-iii} of \cref{app:rulesF} of this article. On the other hand, $\lambda$ denotes the two-particle phase-space function, given as follows:

\begin{align}
    \lambda(s,m_T,m_T)&=\sqrt{\left(1-\frac{2m_T^2}{s}\right)^2-4\left(\frac{m_T^4}{s^2}\right)}.
\end{align}

\subsection{Process $\mu^+\mu^-\to t \bar{T}/\bar{t} T$}

For the single production of the top-quark partner $T$ generated via the  $\mu^+\mu^-\to (\gamma, Z, Z', h, H ) \to t\bar{T}$ process, its corresponding total cross section $ \sigma_{total}(\mu^+\mu^-\to t \bar{T})$  is obtained as follows:

\begin{align} \label{tTtot}
    \sigma_{total}(\mu^+\mu^-\to t \bar{T} )=\sigma_{\gamma}^{t \bar{T} }+\sigma_{Z}^{t \bar{T}}+\sigma_{Z'}^{t \bar{T} } +\sigma_{h}^{t \bar{T}} +\sigma_{H}^{t \bar{T}} +\sigma_{\gamma Z Z' h H}^{t \bar{T} },
\end{align}

\begin{align} \label{tTtot1}
  		\sigma_{\gamma}^{t \bar{T}}=&\frac{(g_V^{\mu\mu\gamma})^2\lambda}{12\pi s^2}\bigg(\left(g_A^{t \bar{T} \gamma}\right)^2\left(\frac{s\lambda^2}{4}-3m_tm_T\right)+\left(g_V^{t \bar{T} \gamma}\right)^2\left(\frac{s\lambda^2}{4}+3m_tm_T\right)\nonumber\\
		&+\frac{3}{4}\left(\left(g_A^{t \bar{T} \gamma}\right)^2+\left(g_V^{t \bar{T} \gamma}\right)^2\right)\sqrt{s\lambda^2+4m_t^2}\sqrt{s\lambda^2+4m_T^2}\bigg),
\end{align}

\begin{align} \label{ww-z}
    \sigma_Z^{t \bar{T}}=&\frac{\lambda\left(\left(g_A^{Z \mu\mu }\right)^2+\left(g_V^{Z \mu\mu }\right)^2\right)}{12\pi\left((s-m_Z^2)^2+m_Z^2\Gamma_Z^2\right)}\bigg(\left(g_A^{t \bar{T} Z}\right)^2\left(\frac{s\lambda^2}{4}-3m_tm_T\right)+\left(g_V^{t \bar{T} Z}\right)^2\nonumber\\
    &\left(\frac{s\lambda^2}{4}+3m_tm_T\right)+\frac{3}{4}\left(\left(g_A^{t \bar{T} Z}\right)^2+\left(g_V^{t \bar{T} Z}\right)^2\right)\sqrt{s\lambda^2+4m_t^2}\sqrt{s\lambda^2+4m_T^2}\bigg),
\end{align}

\begin{align} \label{tTtot2}
   	\sigma_{Z'}^{t \bar{T}}=&\frac{\lambda\left(\left(g_A^{Z' \mu\mu }\right)^2+\left(g_V^{Z' \mu\mu }\right)^2\right)}{12\pi\left((s-m_{Z'}^2)^2+m_{Z'}^2\Gamma_{Z'}^2\right)}\bigg(\left(g_A^{t \bar{T} Z'}\right)^2\left(\frac{s\lambda^2}{4}-3m_tm_T\right)+\left(g_V^{t \bar{T} Z'}\right)^2\nonumber\\
		&\left(\frac{s\lambda^2}{4}+3m_tm_T\right)+\frac{3}{4}\left(\left(g_A^{t \bar{T} Z'}\right)^2+\left(g_V^{t \bar{T} Z'}\right)^2\right)\sqrt{s\lambda^2+4m_t^2}\sqrt{s\lambda^2+4m_T^2}\bigg),
\end{align}

\begin{align} \label{tTtot3x}
    \sigma_h^{t \bar{T} } =&-\frac{g_{h\mu\mu}^2\lambda}{32\pi((s-m_h^2)^2+m_h^2\Gamma_h^2)}\Big(\left(g_A^{t \bar{T} h}\right)^2\left(\sqrt{4m_t^2+\lambda^2s}\sqrt{4m_T^2+\lambda^2s}+4m_tm_T+\lambda^2s\right) \nonumber\\
&+ \left(g_V^{t \bar{T} h}\right)^2\left(\sqrt{4m_t^2+\lambda^2s}\sqrt{4m_T^2+\lambda^2s}-4m_tm_T+\lambda^2s\right)\Big),
\end{align}

\begin{align} \label{tTtot4}
     \sigma_H^{t \bar{T}} =&-\frac{g_{H\mu\mu}^2\lambda}{32\pi((s-m_H^2)^2+m_H^2\Gamma_H^2)}\Big(\left(g_A^{t \bar{T}H}\right)^2\left(\sqrt{4m_t^2+\lambda^2s}\sqrt{4m_T^2+\lambda^2s}+4m_tm_T+\lambda^2s\right)\nonumber\\
&+ \left(g_V^{t \bar{T} H}\right)^2\left(\sqrt{4m_t^2+\lambda^2s}\sqrt{4m_T^2+\lambda^2s}-4m_tm_T+\lambda^2s\right)\Big),
\end{align}

\begin{align} \label{ww-inter}
		\sigma_{\gamma ZZ^\prime h H}^{t\bar{T}}=&
		\frac{\lambda  (g_A^{Z\mu\mu} g_A^{Z'\mu\mu}+g_V^{Z\mu\mu} g_V^{Z'\mu\mu})
			\left(\left(s-m_Z^2\right)
			\left(s-m_{Z'}^2\right)+\Gamma_{Z} \Gamma_{Z'} m_Z m_{Z'}\right) }{24 \pi  \left(\Gamma_{Z}^2
			m_{Z}^2+\left(s-m_{Z}^2\right)^2\right)
			\left(\Gamma_{Z'}^2 m_{Z'}^2+\left(s - m_{Z'}^2\right)^2\right)}\nonumber\\
		&\times \left(F_1 g_V^{t \bar{T} Z} g_V^{t \bar{T} Z'}+F_2 g_A^{t \bar{T} Z} g_A^{t \bar{T} Z'} \right) \nonumber\\
		&+\frac{g_{h\mu\mu} g_{H\mu\mu}
			\lambda
			\left(\left(m_{h}^2-s\right)
			\left(m_{H}^2-s\right)+\Gamma_h \Gamma_H m_{h}
			m_{H}\right) (F_3 g_V^{t\bar{T} h} g_V^{t\bar{T} H}+F_4 g_A^{t \bar{T} h} g_A^{t\bar{T} H})}{16 \pi
			\left(\Gamma_{h}^2 m_{h}^2+\left(s- m_{h}^2\right
			)^2\right) \left(\Gamma_{H}^2 m_{H}^2+\left(s- m_{H}^2\right
			)^2\right)}\nonumber\\
		&+ \frac{g_V^{\mu\mu \gamma} g_{V}^{Z'\mu\mu}
			\lambda  \left(s- m_{Z'}^2\right)(F_{1} g_V^{t \bar{T} \gamma} g_V^{t \bar{T} Z'}+F_{2} g_A^{t \bar{T} \gamma}
			g_A^{t\bar{T}Z'})}{24 \pi  s
			\left(\Gamma_{Z'}^2 m_{Z'}^2+\left(s-m_{Z'}^2\right)^2\right)}\nonumber\\
		&+\frac{g_V^{\mu\mu \gamma} g_{V}^{Z\mu\mu}\lambda  \left(s-m_{Z}^2\right)
			(F_{1} g_V^{t\bar{T}\gamma} g_V^{t\bar{T}Z}+F_{2} g_A^{t\bar{T}\gamma} g_A^{t\bar{T}Z})}{24 \pi  s
			\left(\Gamma_{Z}^2 m_{Z}^2+\left(s- m_{Z}^2\right)^2\right)},
\end{align}  

\noindent where

\begin{align} \label{tTtot-F14}
F_{1,2}=&\pm12m_tm_T + s\lambda^2 +3\sqrt{4m_t^2 + s\lambda^2}\sqrt{4m_T^2 + s\lambda^2},\nonumber\\
F_{3,4}=&(\mp4m_tm_T + s\lambda^2 +\sqrt{4m_t^2 + s\lambda^2}\sqrt{4m_T^2 + s\lambda^2}),\nonumber\\
\lambda\equiv & \lambda(s,m_T,m_t)=\sqrt{\left(1-\frac{m_T^2}{s}-\frac{m_t^2}{s}\right)^2-4\left(\frac{m_T^2}{s}\right)\left(\frac{m_t^2}{s}\right)}.
\end{align}

\noindent As in the previous case,  $g^{x_i}_{V}$ and $g^{x_i}_{A}$  (with $x_i \equiv t \bar{T}\gamma, t \bar{T} Z, t\bar{T} Z', t \bar{T} h, t \bar{T} H$)  denote the vector and axial-vector couplings, respectively (see~\cref{app:rulesF}).

The other single production mechanism for the $T$ quark at a muon collider arises through the $\mu^+\mu^-\to (\gamma, Z, Z', h, H ) \to \bar{t}T$  process. In this case, the corresponding contributions are  similar to those obtained for the $\mu^+\mu^-\to (\gamma, Z, Z', h, H ) \to t \bar{T}$  process. Consequently, the expressions describing the production cross sections for the $T$ quark and the $\bar{t}$ antiquark can be derived from \cref{tTtot,tTtot1,ww-z,tTtot2,tTtot3x,tTtot4,ww-inter,tTtot-F14} by performing the substitutions $t\to T$ and $\bar T\to  \bar t$.

\section{Numerical results} \label{sec:results}

For the numerical analysis of the decay widths, branching ratios, and production cross sections of the top partner $T$, we briefly discuss some of the parameters involved in our calculations, such as the gauge couplings ($g_{A}, g_{B}$), the energy scales $f$ and $F$, the Yukawa couplings $y_i$ ($i=1,2,3$), the mass of the heavy top-quark partner, the masses of the $Z'$ and $H$ bosons, and the ratio of the vacuum expectation values (VEV) of the two Higgs doublets of the BLHM. Next, we discuss these parameters.

\textbf{Gauge couplings ($g_{A}, g_{B}$)}:
The gauge couplings  $g_{A}$ and  $g_{B}$ are associated with the gauge bosons of the $SU(2)_{LA}$ and $SU(2)_{LB}$ groups, respectively~\cite{Schmaltz:2010ac,Kalyniak:2013eva}. These gauge couplings can be parametrized in terms of the mixing angle $\theta_{g}$ through the relation $\tan \theta_{g}=g_{A}/g_{B}$. For simplicity, in our analysis we assume $\tan \theta_{g}=1$, which implies $g_B= g_A$. The couplings $g_{A}$ and  $g_{B}$ are also related to the electroweak gauge coupling $g$ through the relation $g^{-2}=g_{A}^{-2}+g_{B}^{-2}$.
\vspace{0.1cm}

 \textbf{ Energy scales ($f$, $F$)}:
Due to the structure of the BLHM, the model incorporates two distinct global symmetries that are  broken into diagonal subgroups at different energy scales, $f$ and $F$. As a consequence, the energy scales at which the heavy fermions and heavy gauge bosons acquire their masses are different: the fermions obtain masses at the $f$ scale, whereas the gauge boson masses arise at the scale $F>f$. Therefore, the gauge bosons are heavier than the fermions, allowing them to evade  precision electroweak constraints while simultaneously avoiding fine tuning in the top-quark sector.
Regarding the scale $f$, both lower and upper bounds emerge for this parameter once the experimental constraints on heavy-quark production, as well as fine-tuning constraints on the heavy-quark masses, are taken into account, namely $f\in [700,3000]$ GeV~\cite{Kalyniak:2013eva,Cruz-Albaro:2024vjk,Godfrey:2012tf}. On the other hand, the energy scale $F$ takes sufficiently large values compared to the scale $f$, with the purpose of ensuring that the new gauge bosons are heavier than the new quarks; consequently, $F > 3000$ GeV~\cite{Schmaltz:2010ac,Kalyniak:2013eva}.
\vspace{0.1cm}

 \textbf{ Yukawa couplings $(y_1, y_2, y_3)$}:
The Yukawa couplings ($y_1, y_2, y_3$) are associated with the generation of the masses of the six top-quark partners:  $ T $, $ B $, $ T_5 $, $ T_6 $, $ T^{2/3} $, $ T^{5/3} $. The mass eigenstates of the new heavy quarks were determined in Ref.~\cite{Godfrey:2012tf} under the assumption that $y_2 \neq y_3$; otherwise, the masses of the $T$ and $T_5$ quarks become degenerate at lowest order, requiring a different diagonalization scheme. Consequently, for the scenario in which $|y_3-y_2|>0$, the regions of interest are defined as follows:

\begin{itemize}
\item Scenario with $y_2 > y_3$: $y_1=0.61$, $y_2=0.84$, and $y_3=0.35$~\cite{Cruz-Albaro:2023pah,Cruz-Albaro:2022kty,Cruz-Albaro:2022lks},
\item Scenario with $y_2 < y_3$: $y_1=0.61$, $y_2=0.35$, and $y_3=0.84$~\cite{Cruz-Albaro:2023pah,Cruz-Albaro:2022kty,Cruz-Albaro:2022lks}.
\end{itemize}  
\vspace{0.1cm}

 \textbf{Mass of the top-quark partner $T$}:
The mass of the $T$ quark is determined according to~\cref{mT}, where the input parameters are the Yukawa coupling $y_i$, the symmetry breaking scale $f$, and the $\beta$ parameter:
\begin{align} \label{mT}
  m^{2}_T =& (y^{2}_1 + y^{2}_2)f^2 + \frac{9 y^{2}_1  y^{2}_2  y^{2}_3  v^{2}\sin^{2} (\beta)}{(y^{2}_1 + y^{2}_2) (y^{2}_2 - y^{2}_3)}. 
\end{align}

\begin{figure}[H]
\center
\subfloat[]{\includegraphics[width=7.90cm]{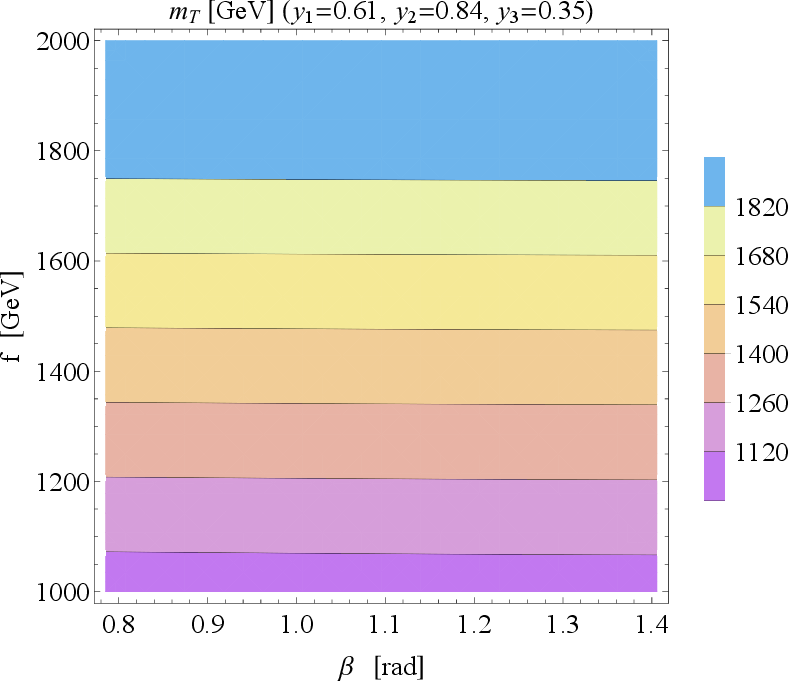}}
\hspace{0.3cm}
\subfloat[]{\includegraphics[width=7.90cm]{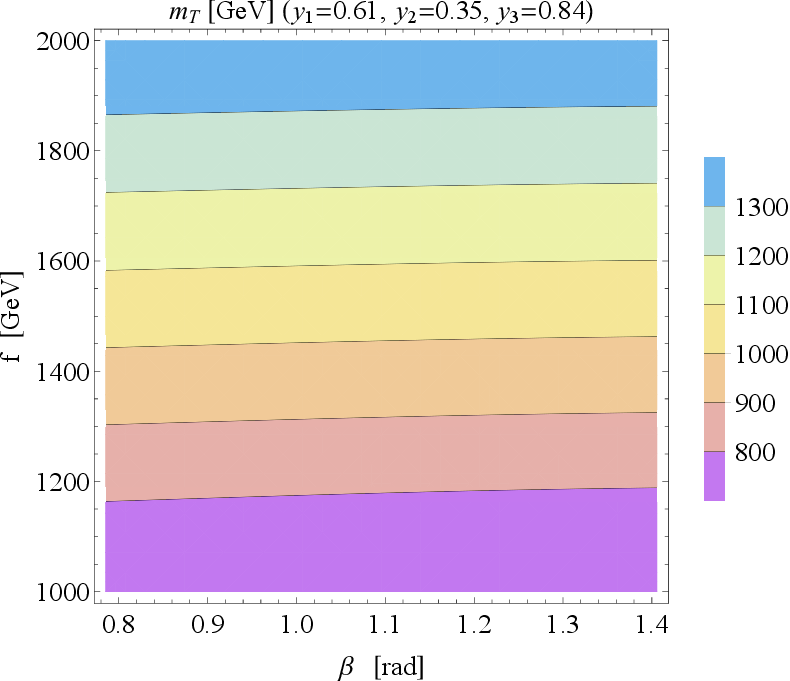} }
\caption{   Contour plots for the mass of the top partner $T$ as a function of the parameters $f\in [1000,2000]$ GeV and $\beta \in [\tan^{-1}(1.0), \tan^{-1}(6.0)] $ rad. The contour plots were generated for the two regions of interest defined by the Yukawa couplings, namely the scenarios $y_2 > y_3$ and $y_2 < y_3$:
a)  $y_1=0.61,\, y_2=0.84,\, y_3=0.35$, and
b)  $y_1=0.61,\, y_2=0.35,\, y_3=0.84$.}
\label{sigmafbeta}
\end{figure}

\noindent In \cref{sigmafbeta}, we present the contour plots of the function $m_{T}(f,\beta)$, which corresponds to the mass of the heavy quark $T$. For this function, the parameters $f$ and $\beta$ are varied ($f\in [1000,2000]$ GeV and $\beta \in [\tan^{-1}(1.0), \tan^{-1}(6.0)] $ rad), while the Yukawa couplings $(y_1, y_2, y_3)$ are kept fixed. From the contour plots, we find that in the region where the Yukawa couplings satisfy $y_2 > y_3$, the values attained by $m_{T}$ are significantly larger than those obtained in the region where $y_2 < y_3$.
The contour plots also show that $m_{T}$ exhibits a stronger dependence on variations of the energy scale $f$. In contrast, the dependence on the parameter $\beta$ is considerably weaker, since $m_{T}$ presents only a mild increase as $\beta$ takes progressively larger values within the proposed interval. Based on these results, and motivated by the fact that current experimental searches for the $T$-quark mass are primarily focused on masses above 1400 GeV, the Yukawa couplings adopted in this study are $y_1=0.61,\, y_2=0.84$, and $ y_3=0.35$.
\vspace{0.1cm}

\textbf{Mass of the $Z'$ gauge boson}:
Within the framework of the BLHM, the new $Z'$ gauge boson acquires a large squared mass proportional to $f^{2}+F^{2}$, as shown in the following equation.
\begin{eqnarray}
m^{2}_{Z'}&=&m^{2}_{W'^{\pm}} +  \frac{g^2 v^4 \sin^{2}\theta_{W} }{16 \cos^{2}\theta_{W} (f^2+F^2)} \left(\sin^{2}\theta_{g} -\cos ^{2} \theta_{g} \right)^{2}, \label{mzprima} \\
m^{2}_{W'^{\pm}}&=& \frac{g^2}{4 \cos^{2}\theta_{g} \sin^{2}\theta_{g}} \left(f^2+F^2 \right)  - m^{2}_{W^{\pm}}, \label{mwprima}
\end{eqnarray}

\noindent where $\theta_{W}$ denotes the Weinberg angle.
\vspace{0.1cm}

  \textbf{Mass of the neutral Higgs boson  $H$}: 
 The mass of the Higgs boson $H$ is calculated from~\cref{massH0}~\cite{Kalyniak:2013eva}: 
 
 \begin{align}\label{massH0}
m^{2}_{H} =& \frac{B_\mu}{\text{sin}\, 2\beta}+ \sqrt{\frac{B^{2}_{\mu}}{\text{sin}^{2}\, 2\beta} -2\lambda_0 B_\mu v^{2} \text{sin}\, 2\beta +\lambda^{2}_{0} v^{4} \text{sin}^{2}\, 2\beta  },
\end{align}

\noindent where

 \begin{eqnarray}
B_\mu &=&\frac{1}{2}(\lambda_0  v^{2} + m^{2}_{A_{0}}  )\, \text{sin}\, 2\beta,\\
\lambda_0 &=& \frac{m^{2}_{h}}{v^{2}}\Big(\frac{  m^{2}_{h}- m^{2}_{A_{0}} }{m^{2}_{h}-m^{2}_{A_{0}} \text{sin}^{2}\, 2\beta }\Big).
\end{eqnarray}

\noindent In these equations, $m_{A_0}$ is the mass of the pseudoscalar $A_0$, and the angle $\beta$ is determined from $\tan\, \beta$, which denotes the ratio of the VEV of the two Higgs doublets in the BLHM.
\vspace{0.1cm}

  \textbf{Ratio of the VEV of the two Higgs doublets for the BLHM ($\tan \beta$)}:
Lower and upper bounds are imposed on the parameter $\tan \beta$. These bounds arise from the perturbativity requirements on the parameter $\lambda_0$ and from the contributions of heavy fermion loops to the Higgs potential~\cite{Schmaltz:2010ac,Kalyniak:2013eva}. Consequently, the allowed range of values for $\tan \beta$ is determined according to the following equation,

 \begin{eqnarray}\label{parametros}
1 < &\text{tan}\, \beta &  <  \sqrt{ \frac{2+2 \sqrt{\big(1-\frac{m^{2}_{h} }{m^{2}_{A_0}} \big) \big(1-\frac{m^{2}_{h} }{4 \pi v^{2}}\big) } }{ \frac{m^{2}_{h}}{m^{2}_{A_0}} \big(1+ \frac{m^{2}_{A_0}- m^{2}_{h}}{4 \pi v^{2}}  \big) } -1 }.
\end{eqnarray}

\noindent The upper bound on the parameter $\tan \beta$ is primarily determined by the input parameter $m_{A_0}$. In our analysis, we consider values consistent with current experimental searches for the pseudoscalar Higgs boson $A_0$, imposing the constraint $m_{A_0}\geq 500$ GeV~\cite{ATLAS:2020gxx,CMS:2019ogx}.
\vspace{0.1cm}

 \textbf{ Total decay widths}:
The total decay widths $\Gamma_h$, $\Gamma_Z$, $\Gamma_H$, and $ \Gamma_{Z'}$ appearing in~\cref{width-z,width-z1,width-z2,width-z3,TT2} and~\cref{ww-z,tTtot2,tTtot3x,tTtot4,ww-inter} are treated as follows. The decay widths $\Gamma_h$ and $\Gamma_Z$ are experimentally measured, and their numerical values are listed in~\cref{parametervalues}. In contrast, the decay widths $\Gamma_H$ and $\Gamma_{Z'}$ are taken from the calculations reported in Refs.~\cite{Cruz-Albaro:2024vjk,Gutierrez-Rodriguez:2023sxg}.
\vspace{0.1cm}

 \cref{parametervalues} lists the values assigned to the input parameters involved in our calculation of single and pair production of the top partner $T$ at the future muon collider within the framework of the BLHM.

\begin{table}[H]
\caption{Values assigned to the parameters involved in our calculation of top partner $T$ production at the muon collider.
\label{parametervalues}}
\centering
\begin{tabular}{|c | c | c |}
\hline
\hspace{0.5cm} $ \textbf{Parameter} $ \hspace{0.5cm}  &  \hspace{1.2cm}  $\textbf{Value} $ \hspace{1.2cm}  &  \hspace{0.5cm}   $ \textbf{Reference} $ \hspace{0.5cm} \\
\hline
\hline
$ m_{h}  $  &   $ 125.25\  \text{GeV} $ &  \cite{pdglive}  \\
\hline
$ g_{B}=g_{A} $  &   $ \sqrt{2}\, g $ & \cite{Cruz-Albaro:2023pah,Cruz-Albaro:2022kty,Cruz-Albaro:2022lks}  \\
\hline
$ f $  &  $ [1000, 3000]\   \text{GeV} $ &  \cite{Schmaltz:2010ac,Cruz-Albaro:2023pah, Cruz-Albaro:2022kty,Cruz-Albaro:2022lks,Kalyniak:2013eva}  \\
\hline
$ \tan \beta $  &    $ [1,6]  $  &  \cite{Cruz-Albaro:2023pah,Cruz-Albaro:2022kty,Cruz-Albaro:2022lks} \\
\hline
$ (y_1, y_2, y_3) $  &   $ (0.61, 0.84, 0.35) $ &  \cite{Cruz-Albaro:2023pah,Cruz-Albaro:2022kty,Cruz-Albaro:2022lks}  \\
\hline
$ \Gamma_{h} $  &   $ 3.00 \times 10^{-3} $ GeV &  \cite{pdglive}   \\
\hline
$ \Gamma_{Z} $  &   $ 2.49 $ GeV &  \cite{pdglive}   \\
\hline
\end{tabular}
\end{table}

\subsection{Decay widths  of the top partner $T$}

In this subsection, we discuss the decay widths $\Gamma(T \to X)$ of the top partner $T$ as a function of the energy scale $f$, the parameter $\beta$, and the mass of the  $T$ quark ($m_{T}$), where $T \to X$ denotes the leading tree-level decay channels of this new heavy quark, namely $X \equiv t h$, $t \gamma$, $tZ$, $W b$, $tZh$, $hWb$, $ZWb $, $t WW$, $tZZ$.

\begin{figure}[H]
\center
\subfloat[]{\includegraphics[width=8.0cm]{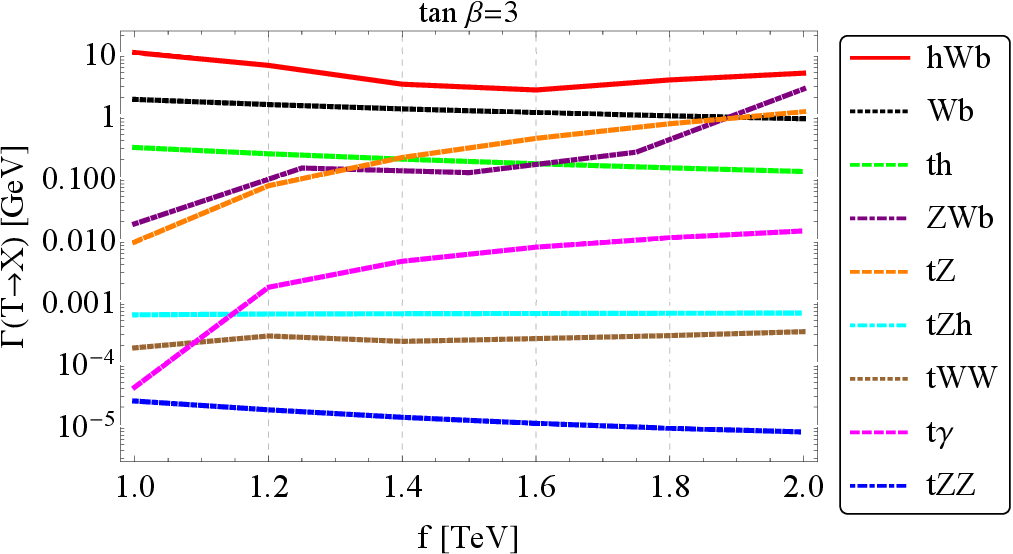}}  \hspace{0.3cm}
\subfloat[]{\includegraphics[width=8.0cm]{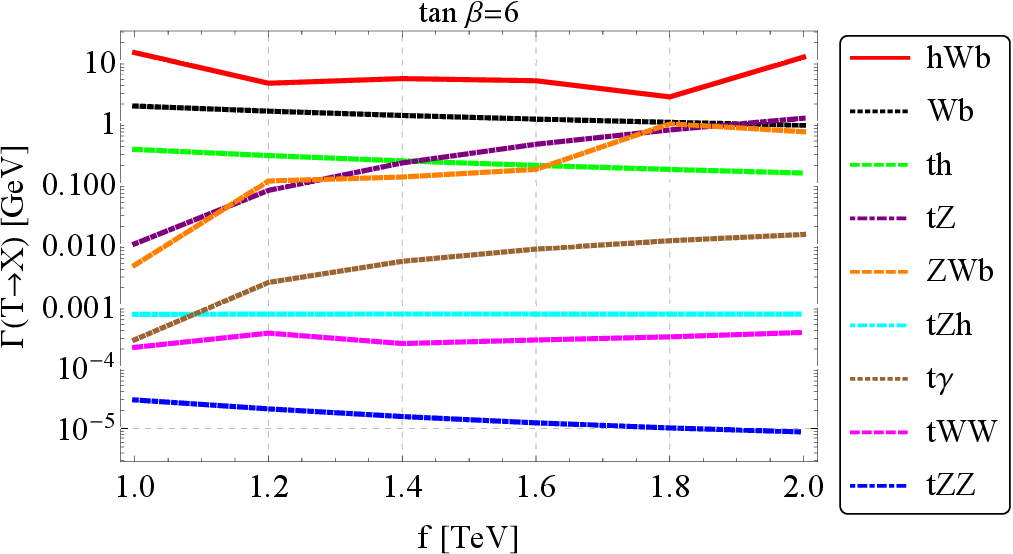}}
\caption{ Decay widths for the $T\to X$ processes (where $X \equiv t h$, $t \gamma$, $tZ$, $W b$, $tZh$, $hWb$, $ZWb $, $t WW$, $tZZ$) as a function of the energy scale $f$, for two fixed values of the parameter $\beta$:  (a) $\tan \beta=3$, (b) $\tan \beta=6$.} \label{width-f}
\end{figure}

We begin by discussing the behavior of $\Gamma(T \to X)$ where the new physics scale $f$ varies from 1000 to 2000 GeV, while the parameter $\beta$ is fixed at two benchmark values, $\tan \beta= 3$ and $\tan \beta=6$.
 As shown in  \cref{width-f}(a), for  $\tan \beta= 3$, the decay width for the $T \to hWb$ channel represented by the curve $\Gamma(T \to hWb)$ provides the dominant contribution throughout the analyzed $f$ range. In contrast, the smallest contribution is associated with the curve $\Gamma(T \to tZZ)$.
 The numerical contributions provided by these two curves are $\Gamma(T \to hWb)=[10.53, 4.90]$ GeV and $\Gamma(T \to tZZ)=[2.50\times 10^{-5}, 7.93 \times 10^{-6}]$ GeV, respectively.
 We also note that the subdominant contribution arises from the $T\to Wb$ decay channel, whose decay width is $\Gamma(T \to Wb) = [1.85, 1.13 ]$ GeV while $f\approx [1000,1600]$ GeV.  
 In~\cref{width-f}(b), generated for $\tan \beta= 6$, we find once again that the channels yielding the dominant and most suppressed decay widths are $T \to hWb$ and $T \to tZZ$, respectively. The corresponding decay widths are $\Gamma(T \to hWb)=[13.87, 11.69]$ GeV and $\Gamma(T \to tZZ)=[2.93\times 10^{-5}, 8.79 \times 10^{-6}]$ GeV. On the other hand, the subdominant contribution is provided by the two-body decay channel $T \to Wb$, represented by the decay width $\Gamma(T \to Wb)$, where $\Gamma(T \to Wb)=[1.86,1.01]$ GeV when $f\approx [1000,1800]$ GeV.
 Based on the above discussion and the results displayed in~\cref{width-f}(a) and~\cref{width-f}(b), we conclude that the function $\Gamma(T \to X)$ exhibits a strong dependence on the symmetry breaking scale $f$.  This behavior is evident from the corresponding curves, which undergo significant variations as $f$ increases within the allowed parameter space. Furthermore, for the two benchmark scenarios considered for the parameter $\beta$, we find that the three-body decay channel $T \to hWb$ provides the dominant contribution to the  decay width of the top partner $T$.

\begin{figure}[H]
  \center
  \subfloat[]{\includegraphics[width=8.0cm]{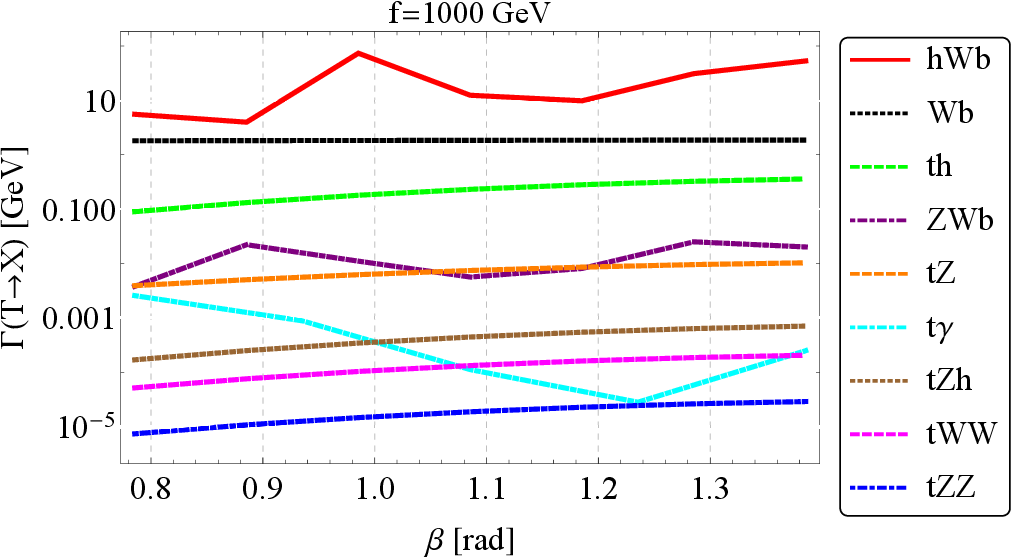}}  \hspace{0.3cm}
\subfloat[]{\includegraphics[width=8.0cm]{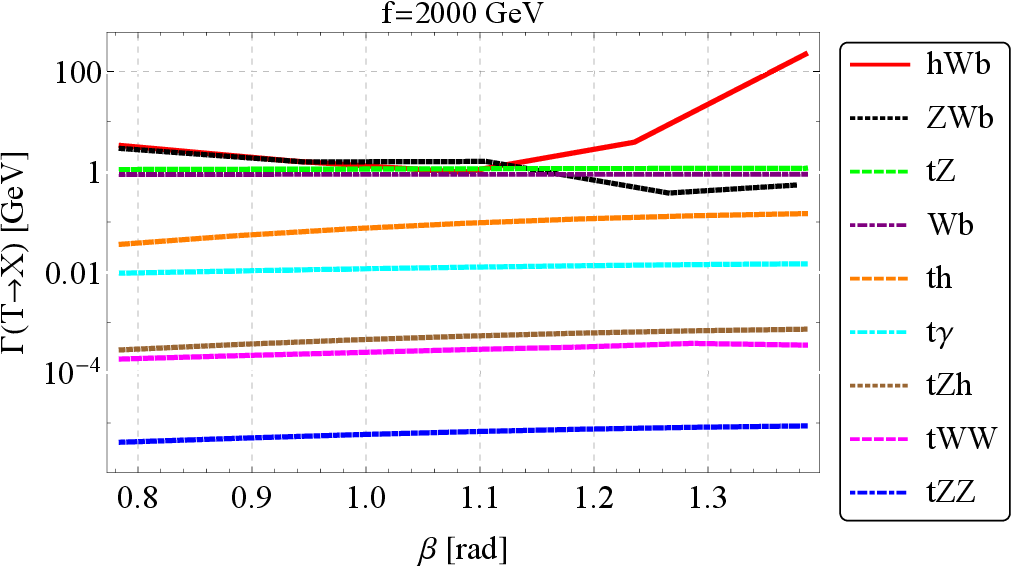}}
  \caption{Decay widths for the $T\to X$ processes (where $X \equiv t h$, $t \gamma$, $tZ$, $W b$, $tZh$, $hWb$, $ZWb $, $t WW$, $tZZ$) as a function of the parameter $\beta$, for two fixed values of the energy scale $f$:  (a) $f=1000$ GeV, (b) $f=2000$ GeV.}\label{width-beta}
\end{figure}

We now turn to a discussion of the dependence of $\Gamma(T \to X)$ on the parameter $\beta$ as depicted in~\cref{width-beta}.
For this scenario, the plots are obtained by setting $f = 1000$ GeV and $f = 2000$ GeV when $\beta \in [\tan^{-1}(1.0), \tan^{-1}(6.0)]$ rad.
In the left plot of~\cref{width-beta}, it can be observed that the dominant and subdominant contributions to the decay width of the heavy quark $T$ arise from the tree-level three-body and two-body decay channels, $T \to hWb$ and $T \to Wb$, respectively. This hierarchy remains unchanged throughout the entire range of the parameter $\beta$ considered in our analysis.  The corresponding numerical contributions associated with these decays are $\Gamma(T \to hWb)=[5.53, 58.89]$ GeV and $\Gamma(T \to Wb)=[1.81, 1.86]$ GeV, respectively. On the other hand, $\Gamma(T \to tZZ)$ and $\Gamma(T \to t\gamma)$  give contributions reaching numerical values up to the order $10^{-6}$ GeV. 
Concerning the right plot of~\cref{width-beta}, in this case, the decays associated with $T \to hWb$ and $T \to ZWb$
provide the leading contributions to  $\Gamma(T \to X)$, as evidenced by the corresponding curves, which exhibit a comparable behavior and appear to compete with one another in the range of $\beta \approx [0.78, 1.14]$ rad: $\Gamma(T \to hWb)=[3.33, 1.26]$ GeV and $\Gamma(T \to ZWb)=[2.91, 1.61]$ GeV. Beyond this region, however, the dominant contribution is provided by the $T \to hWb$ decay channel with  $\Gamma(T \to hWb)=[1.26, 219.68]$ GeV when $\beta \approx [1.14, 1.40]$ rad.  Conversely, the weakest contribution comes from $\Gamma(T \to tZZ)$, which contributes on the order of $10^{-6}$ GeV. 
The analysis of the dependence of $\Gamma(T \to X)$ versus $\beta$ indicates that  $\beta$ has only a small impact on the decay widths of certain decay channels of the top partner $T$, such as $T\to Wb$ and $T\to tZh$. In contrast, other channels, such as the three-body decay $T \to hWb$, exhibit a significantly stronger sensitivity to variations of the parameter $\beta$.

\begin{figure}[H]
\center
{\includegraphics[width=8.0cm]{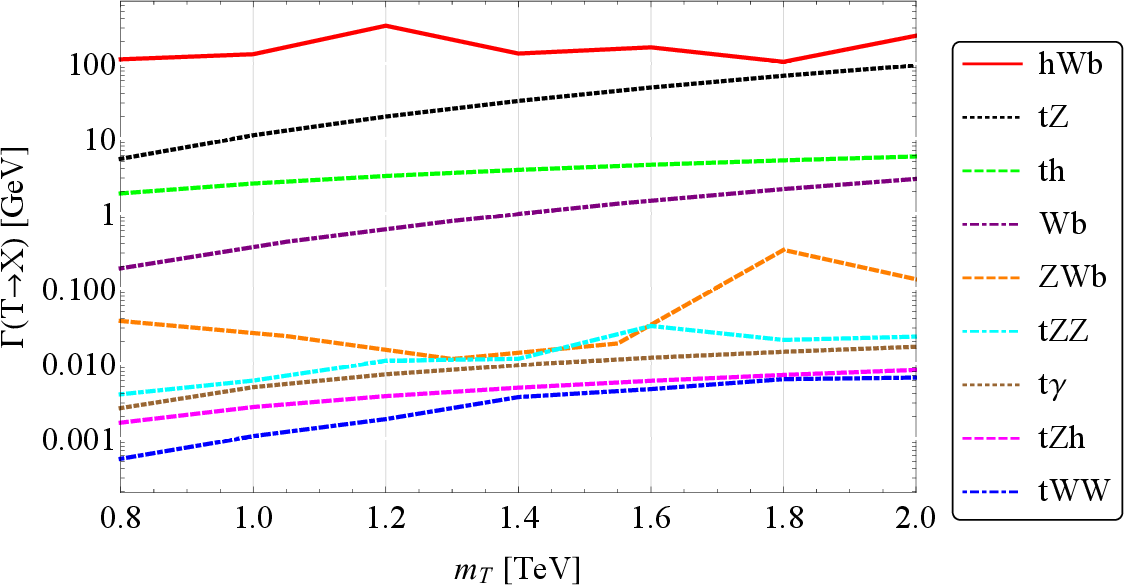}} 
\caption{Decay widths for the $T\to X$ processes (where $X \equiv t h$, $t \gamma$, $tZ$, $W b$, $tZh$, $hWb$, $ZWb $, $t WW$, $tZZ$) as a function of  $m_{T}$.} \label{width-mT} 
\end{figure}

Finally, we investigate the dependence of the decay width $\Gamma(T \to X)$ on the mass of the top partner $T$, as illustrated in~\cref{width-mT}. As in the previously discussed scenarios, the numerical analysis shows that the tree-level decay channel $T\rightarrow h W b$ provides the dominant contribution to $\Gamma(T \to X)$ over the entire mass range considered. In contrast, the three-body tree-level decay $T\rightarrow tWW$ yields the smallest contribution. The corresponding numerical values for these decay channels are  $\Gamma(T \to h W b)=[114.69, 236.70]$ GeV and  $\Gamma(T \to t W W)=[5.43 \times 10^{-4}, 6.56 \times 10^{-3}]$ GeV, respectively, for $m_{T}\in[800,2000]$ GeV. 
It is worth emphasizing that all curves shown in~\cref{width-mT} have been obtained for the benchmark values $f=1000$ GeV and $\tan \beta=3$. Nevertheless, varying these parameters within the ranges considered in this work leads to qualitatively similar behaviors. 

\subsection{Branching ratios of the top partner $T$}

\begin{figure}[H]
\center
\subfloat[]{\includegraphics[width=8.0cm]{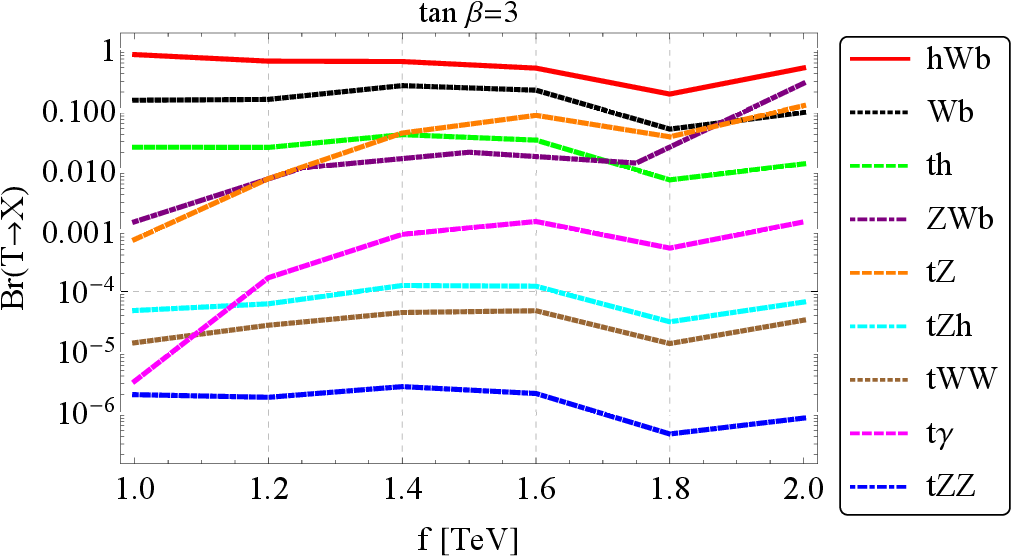}} \hspace{0.3cm}
\subfloat[]{\includegraphics[width=8.0cm]{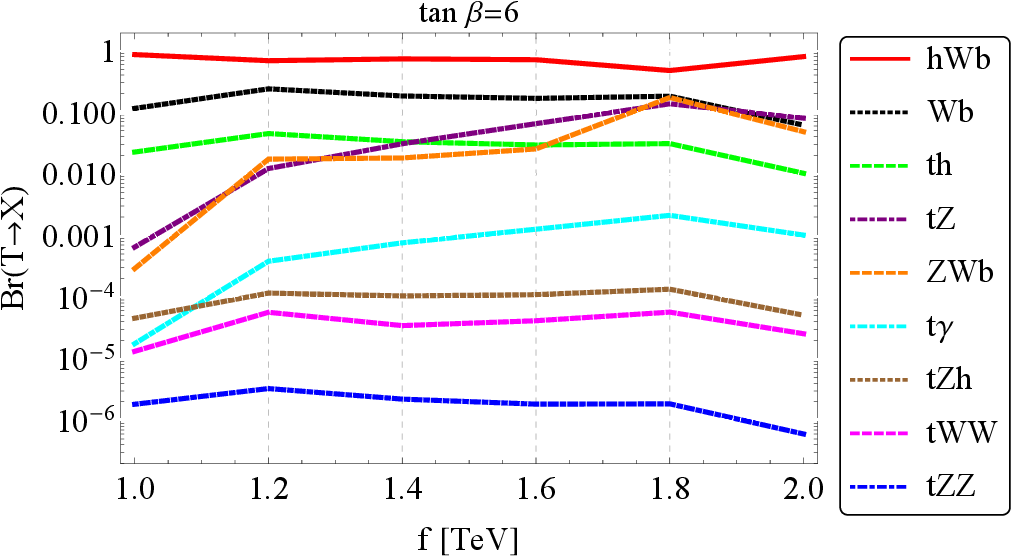}}
\caption{ Branching ratios for the $T\to X$ processes (where $X \equiv t h$, $t \gamma$, $tZ$, $W b$, $tZh$, $hWb$, $ZWb $, $t WW$, $tZZ$) as a function of the energy scale $f$, for two fixed values of the parameter $\beta$:  (a) $\tan \beta=3$,  (b) $\tan \beta=6$.} \label{branching-f}
\end{figure}

\begin{figure}[H]
\center
\subfloat[]{\includegraphics[width=8.0cm]{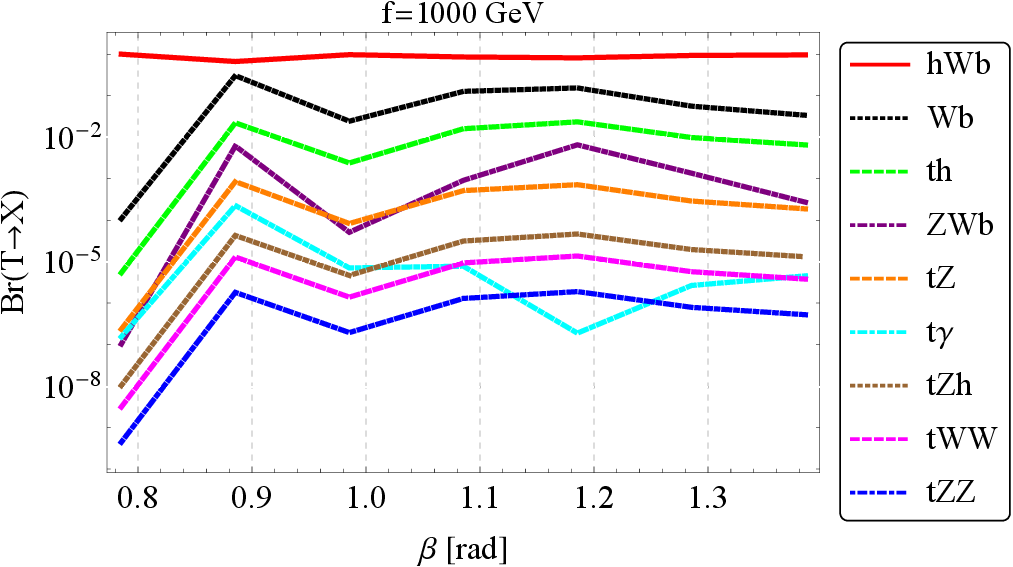}} \hspace{0.3cm}
\subfloat[]{\includegraphics[width=8.0cm]{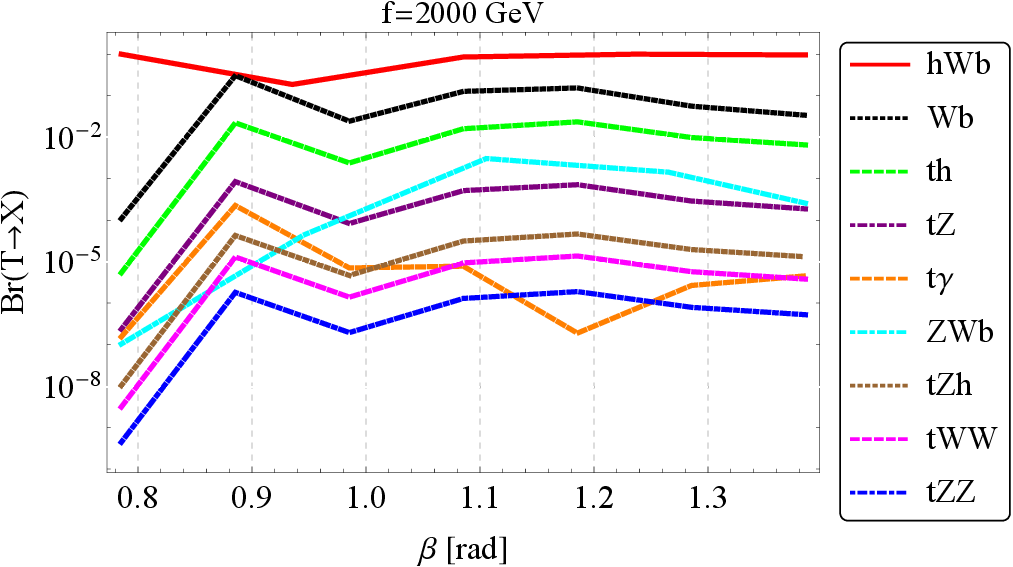}}
\caption{ Branching ratios for the $T\to X$ processes (where $X \equiv t h$, $t \gamma$, $tZ$, $W b$, $tZh$, $hWb$, $ZWb $, $t WW$, $tZZ$) as a function of the parameter $\beta$, for two fixed values of the energy scale $f$:  (a) $f=1000$ GeV,  (b) $f=2000$ GeV.} \label{branching-beta}
\end{figure}

\begin{figure}[H]
\center
{\includegraphics[width=8.0cm]{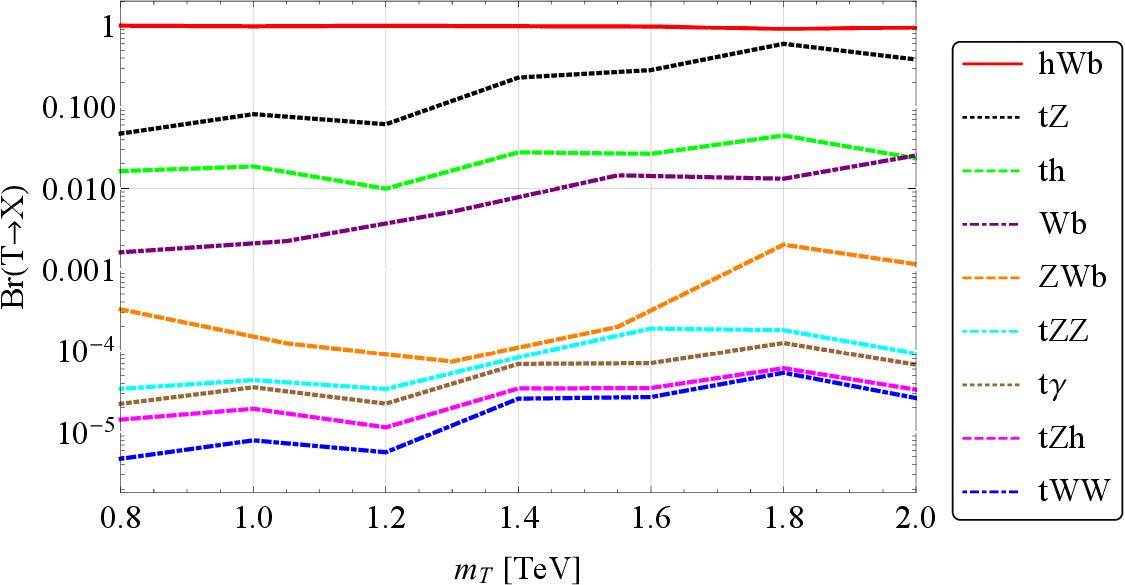}}
\caption{Branching ratios for the $T\to X$ processes (where $X \equiv t h$, $t \gamma$, $tZ$, $W b$, $tZh$, $hWb$, $ZWb $, $t WW$, $tZZ$) as a function of $m_{T}$.} \label{branching-mT} 
\end{figure}

The branching ratios provide essential information regarding the probabilities that the top partner $T$ decays into a specific final state. It is important to emphasize that the determination of these branching ratios requires the calculation of the total decay width of the $T$ quark; for this purpose, all relevant decay channels of the particle under consideration have been taken into account, i.e., $T \to t h$, $t \gamma$, $tZ$, $W b$, $tZh$, $hWb$, $ZWb $, $t WW$, $tZZ$. Following the approach adopted in the previous subsection, we now analyze the branching ratios of the heavy quark $T$, $\text{Br}(T \to X)$, as functions of the new-physics scale $f$, the parameter $\beta$, and the mass of the  $T$ quark  as illustrated in \cref{branching-f,branching-beta,branching-mT}, respectively. 
We first examine the behavior of  $\text{Br}(T \to X)$ vs. $f$ by setting two fixed scenarios for the parameter $\tan \beta$ ($\tan \beta=3$ and $\tan \beta=6$), as shown in \cref{branching-f}(a) and~\cref{branching-f}(b). From~\cref{branching-f}(a), the numerical evaluation shows that for values of $f \approx [1000, 1880]$ GeV, the magnitudes of the branching ratios that generate dominant and subdominant contributions are $\text{Br}(T \to hWb)=[8.28, 1.85]\times 10^{-1}$ and $\text{Br}(T \to Wb)=[1.45 \times 10^{-1}, 4.85 \times 10^{-2}]$. On the other hand, for $\tan \beta=6$, we find that the tree-level decay channels $T \to hWb$ and $ T \to Wb$ also provide the dominant and subdominant branching ratios over nearly the entire $f$ interval considered, as illustrated in \cref{branching-f}(b). In this case, the corresponding branching ratios turn out to be  $\text{Br}(T \to hWb)=[8.61, 7.98]\times 10^{-1}$ and $\text{Br}(T \to Wb)=[1.15 \times 10^{-1}, 6.17 \times 10^{-2}]$.
It is worth noting that for both benchmark scenarios considered for  $\tan \beta$, the decay channel $T\to tZZ$ yields branching ratios of the order of $10^{-6}$ to $10^{-7}$ for $f\in [1000,2000]$ GeV.
These results clearly indicate that the branching ratio  $\text{Br}(T \to X)$ exhibits a strong dependence on the new-physics scale $f$.

As far as the behavior of $\text{Br}(T \to X)$  as a function of $\beta$ is concerned, we observe in \cref{branching-beta}(a) that when $f = 1000$ GeV, the most probable decay channels for the heavy quark $T$ are given by the processes $T\to hWb$ and $T \to Wb$ when $\beta \in [\tan^{-1}(1.0), \tan^{-1}(6.0)]$ rad. These decays arise at the tree level and generate dominant and subdominant contributions to $\text{Br}(T \to X)$: 
$\text{Br}(T \to hWb)=[9.98, 9.59]\times 10^{-1}$ and $\text{Br}(T \to Wb)=[1.08 \times 10^{-4}, 3.37 \times 10^{-2}]$, respectively.
On the other hand, it can also be seen that the $T\to tZZ$  decay generates small contributions to the branching ratio, that is, $\text{Br}(T \to tZZ)=[4.44\times 10^{-10}, 5.25\times 10^{-7}] $.
Regarding the plot shown in \cref{branching-beta}(b), which was obtained by fixing the parameter $f$ at 2000 GeV, the curves denoted by $\text{Br}(T \to hWb)=[9.99, 9.59]\times 10^{-1}$ and $\text{Br}(T \to Wb)=[1.08 \times 10^{-4}, 3.37 \times 10^{-2}]$ provide the dominant contributions throughout the entire $\beta$ interval considered in our analysis. In contrast, the smallest contribution in this scenario comes from the decay channel  $T\to tZZ$, which yields a branching ratio of  $\text{Br}(T \to tZZ)=[4.43\times 10^{-10}, 5.25\times 10^{-7}] $.
In our analysis of  $\text{Br}(T \to X)$  vs. $\beta$, the $T\to hWb$ and $T \to Wb$ decay channels yield the highest probabilities for the particle of interest to decay via these specific channels. Furthermore, $\text{Br}(T \to X)$ shows greater dependence on the $\beta$ parameter; the effects are quite noticeable in the curves (see~\cref{branching-beta}).

The dependence of the branching ratio $\mathrm{Br}(T\rightarrow X)$ on the mass of the top partner $T$ is presented in~\cref{branching-mT}.
 As can be seen in this figure, the tree-level decay channel $T\rightarrow hWb$ provides the dominant branching ratio over the entire mass range considered for $m_{T}$, that is, $m_{T} \in [800, 2000]$ GeV. In this case, the numerical contribution is $\mathrm{Br}(T\rightarrow hWb)=[9.98, 9.50]\times 10^{-1}$.
 By contrast, the three-body decay $T\rightarrow tWW$ yields a significantly suppressed branching ratio  of $10^{-6}$ to $10^{-5}$. 

\subsection{Pair  production cross section of the top partner $T$}
We now present an analysis of the pair-production cross section of the top-quark partner $T$ mediated by gauge ($\gamma, Z, Z'$) and Higgs ($h, H$) boson exchange in muon collisions at the center-of-mass energies envisioned for a future muon collider, which correspond to the center-of-mass energies of $\sqrt{s}=3000$ GeV and $\sqrt{s}=10000$ GeV~\cite{Accettura:2023ked,Long:2020wfp}. Owing to its clean experimental environment and leptonic initial state, a muon collider could probe heavy particles more effectively than current or future hadron colliders, where signal sensitivity is often reduced by overwhelming QCD backgrounds.  Its potential to explore uncharted territory stems from its unique combination of high center-of-mass energy, extreme precision, and suppressed hadronic background.

\begin{figure}[H]
\center
\subfloat[]{\includegraphics[width=8.0cm]{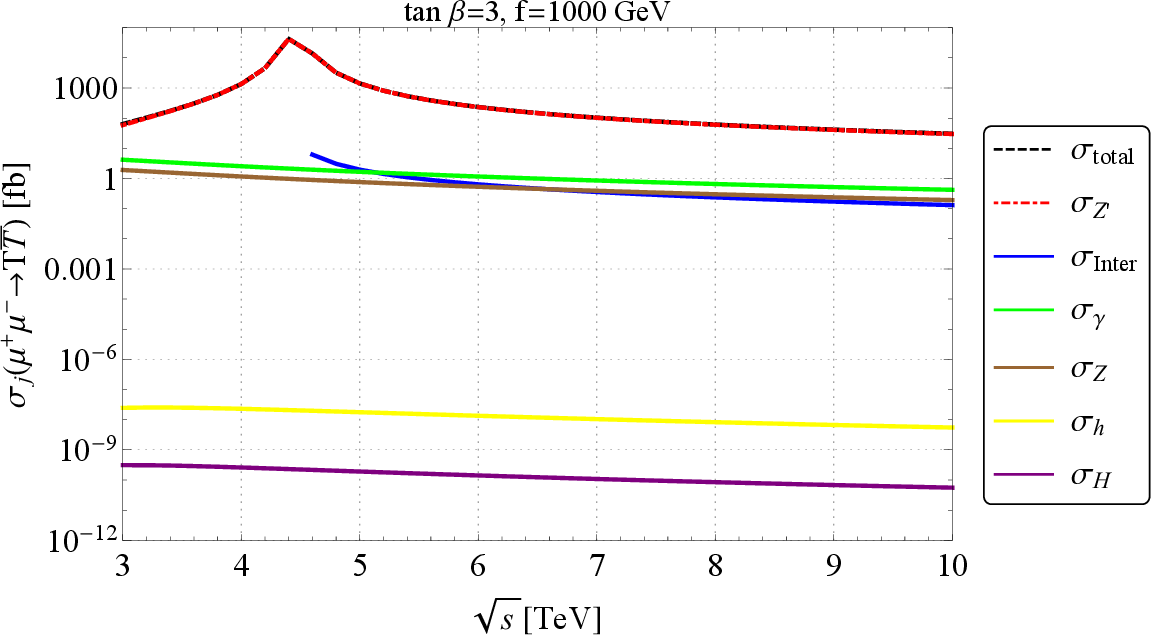}}\vspace{0.6cm}\hspace{0.3cm}
\subfloat[]{\includegraphics[width=8.0cm]{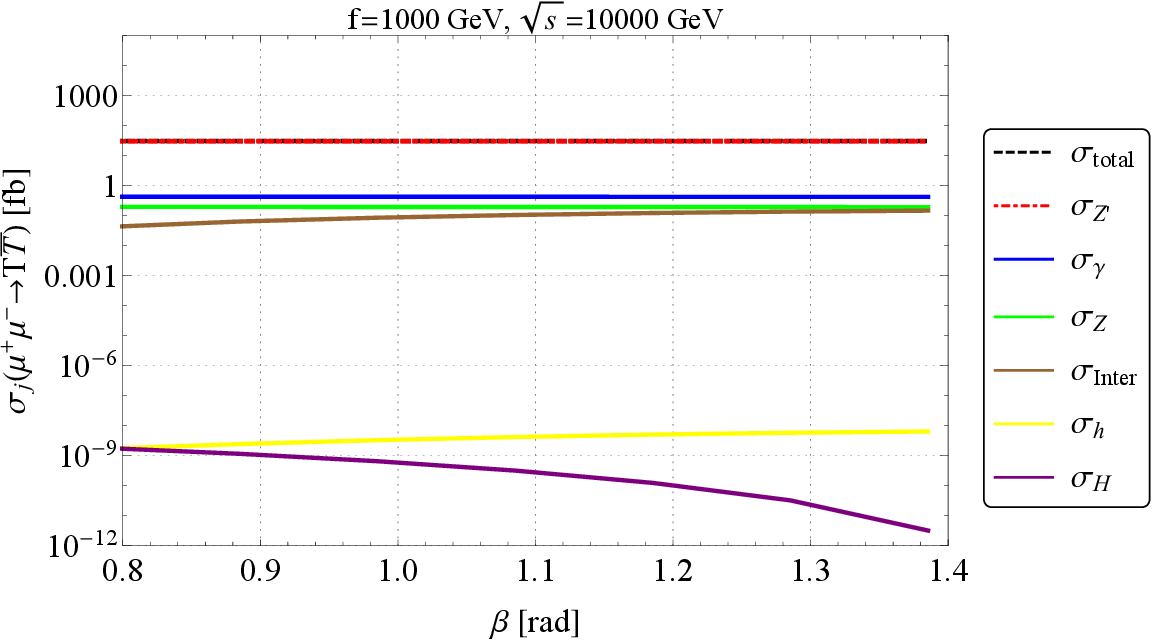}} 
\caption{Total and partial cross sections of the process $\mu^{+} \mu^{-}  \to T\bar{T}$: (a) as a function of the center-of-mass energy $\sqrt{s}$ and (b) as a function of the parameter $\beta$.} \label{sigma-partial} 
\end{figure}

In \cref{sigma-partial}, we display the total cross section $\sigma_{total}(\mu^{+} \mu^{-} \to T\bar{T})$ together with its individual contributions. These partial cross sections, denoted by $\sigma_{i}(\mu^{+} \mu^{-} \to T\bar{T})$ (with $i \equiv \gamma, Z, Z', h, H, \gamma Z Z' h H$), arise from the processes $\mu^{+} \mu^{-}\to i \to T\bar{T}$. It should be noted that $\sigma_{\gamma Z Z' h H}(\mu^{+} \mu^{-} \to T\bar{T})$ (where $ \sigma_{\gamma Z Z' h H} \equiv \sigma_{\text{Inter}}$)  denotes the cross section associated with the interference effects between the gauge and Higgs bosons. Thus,  the total cross section $\sigma_{total}(f, \beta, \sqrt{s})$ is determined from the sum of the partial cross sections (see~\cref{tot-TT}).
In the left plot of \cref{sigma-partial}, we present the behavior of the curves corresponding to the partial contributions to $\sigma_{total}(\mu^{+} \mu^{-} \to T\bar{T})$, together with the total cross section, as a function of the center-of-mass energy $\sqrt{s}$ for $\tan\beta = 3$ and $f = 1000$ GeV, and in the right plot we set  $f=1000$ GeV and $\sqrt{s}=10000$ GeV and show the dependence of $\sigma_{total}(\mu^{+} \mu^{-} \to T\bar{T})$ and $\sigma_{i}(\mu^{+} \mu^{-} \to T\bar{T})$ on the parameter $\beta$.
In the first benchmark scenario (see~\cref{sigma-partial}(a)), we find that the dominant contribution to the total cross section over most of the parameter space  is provided by the process $\mu^{+}\mu^{-} \rightarrow Z' \rightarrow T\bar{T}$ when the center-of-mass energy $\sqrt{s}$ varies from 3000 to 10000 GeV. In contrast, the contribution mediated by the new Higgs boson yields the smallest contribution to the cross section $\sigma_{total}(\mu^{+}\mu^{-} \rightarrow T\bar{T})$. As shown in~\cref{sigma-partial}(a), the cross sections $\sigma_{Z'}(\mu^{+}\mu^{-}  \rightarrow T\bar{T})$ and $\sigma_{total}(\mu^{+}\mu^{-} \rightarrow T\bar{T})$ reach their resonance peaks when the center-of-mass energy is $\sqrt{s} \approx 4400$ GeV:   $\sigma_{Z'}(\mu^{+}\mu^{-}  \rightarrow T\bar{T})=42403.50$ fb  and $\sigma_{total}(\mu^{+}\mu^{-} \rightarrow T\bar{T})=42399.00$ fb.  The difference between these values is due to negative interference occurring at this point. It is also worth noting that the cross section associated with the interference among the mediating particles involved in the pair production of the $T$ quark gives rise to constructive interference in the energy range $\sqrt{s}\approx [4600,10000 ]$ GeV. Regarding the second benchmark scenario, namely the dependence of the cross section on the parameter $\beta$, we find that the dominant and suppressed contributions to the total cross section are once again provided by the processes $\mu^{+}\mu^{-} \rightarrow Z' \rightarrow T\bar{T}$ and $\mu^{+}\mu^{-} \rightarrow H \rightarrow T\bar{T}$, respectively, as shown in~\cref{sigma-partial}(b). 
The numerical contributions generated for these decay channels are  $\sigma_{Z'}(\mu^{+}\mu^{-} \rightarrow T\bar{T})=[29.82, 29.83]$ fb and  $\sigma_{H}(\mu^{+}\mu^{-} \rightarrow T\bar{T})=[1.80 \times 10^{-9}, 3.17 \times 10^{-12}]$ fb when  $\beta \in [\tan^{-1}(1.0), \tan^{-1}(6.0)]$ rad; within this same region, $\sigma_{total}(\mu^{+}\mu^{-} \rightarrow T\bar{T})=[30.48,30.58]$ fb is obtained.
It is worth emphasizing that  the interference contribution remains positive throughout the entire $\beta$ range considered, indicating the presence of constructive interference over the full parameter space analyzed.
From the numerical results, we conclude that the cross sections mediated by gauge bosons for the pair production of the top partner $T$ are several orders of magnitude larger than those arising from Higgs bosons exchange. This behavior reflects the dominant role played by gauge interactions in the production mechanism of the heavy quark at  the proposed muon collider.

\begin{figure}[H]
\center
\includegraphics[width=8.0cm]{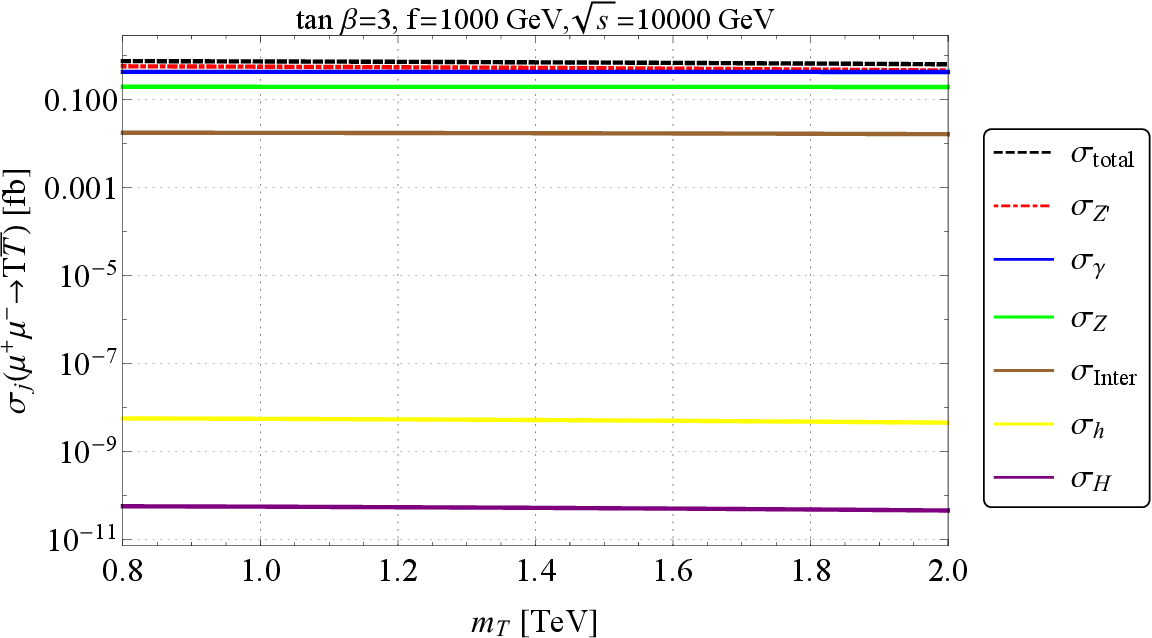}
\caption{Total and partial cross sections of the process $\mu^{+} \mu^{-}  \to T\bar{T}$ as a function of $m_{T}$.} \label{sigma-partial-mT} 
\end{figure}

At this stage, it is also instructive to investigate the dependence of the partial and total cross sections for top partner pair production on the mass of the heavy quark $T$. As shown in \cref{sigma-partial-mT}, the $Z'$ and photon ($\gamma$) mediated channels provide the dominant and subdominant contributions, respectively, to the total production cross section $\sigma_{total}(\mu^{+}\mu^{-} \rightarrow T\bar{T})$ over the mass range $m_{T}\in[800, 2000]$ GeV:  $\sigma_{Z'}(\mu^{+}\mu^{-} \rightarrow T\bar{T})=[5.61, 4.49] \times 10^{-1}$ fb and  $\sigma_{\gamma}(\mu^{+}\mu^{-} \rightarrow T\bar{T})=[4.21, 4.16] \times 10^{-1} $ fb. In contrast, the contributions arising from the Higgs-mediated channels remain strongly suppressed throughout the entire region of parameter space considered:  $\sigma_{h}(\mu^{+}\mu^{-} \rightarrow T\bar{T})=[5.66,4.53] \times 10^{-9}$ fb and  $\sigma_{H}(\mu^{+}\mu^{-} \rightarrow T\bar{T})=[5.72, 4.58] \times 10^{-11} $ fb. Overall, the partial production cross sections exhibit a weak dependence on the top partner mass. 

\begin{figure}[H]
\center
\subfloat[]{\includegraphics[width=8.0cm]{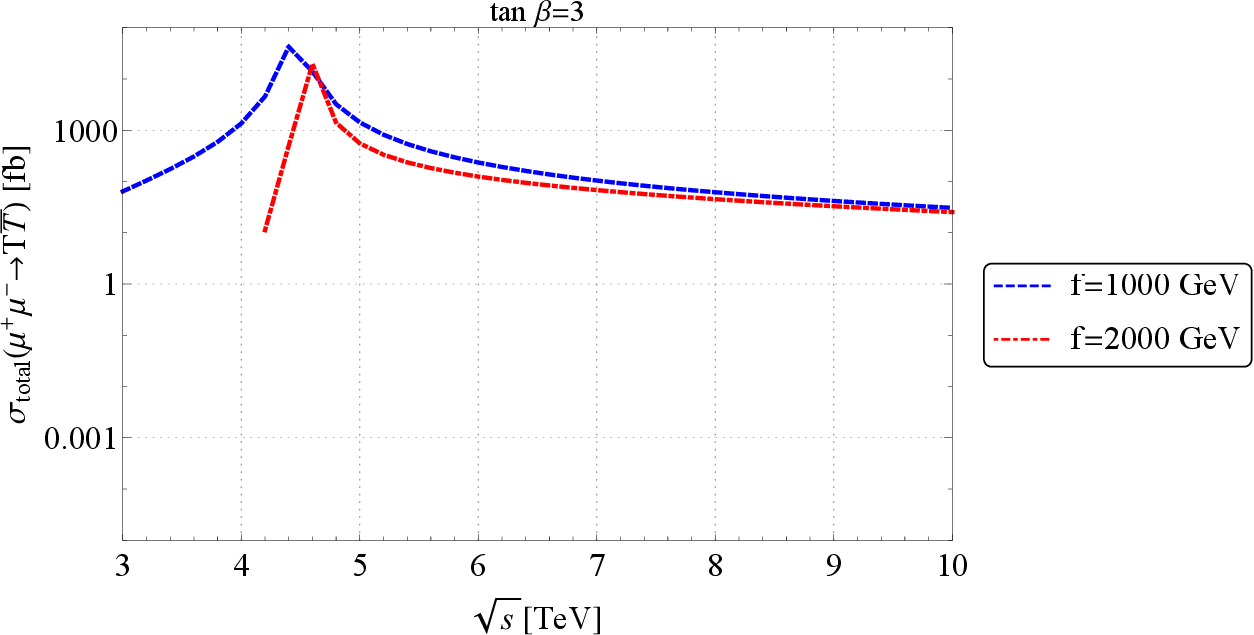}}\vspace{0.6cm}\hspace{0.3cm}
\subfloat[]{\includegraphics[width=8.0cm]{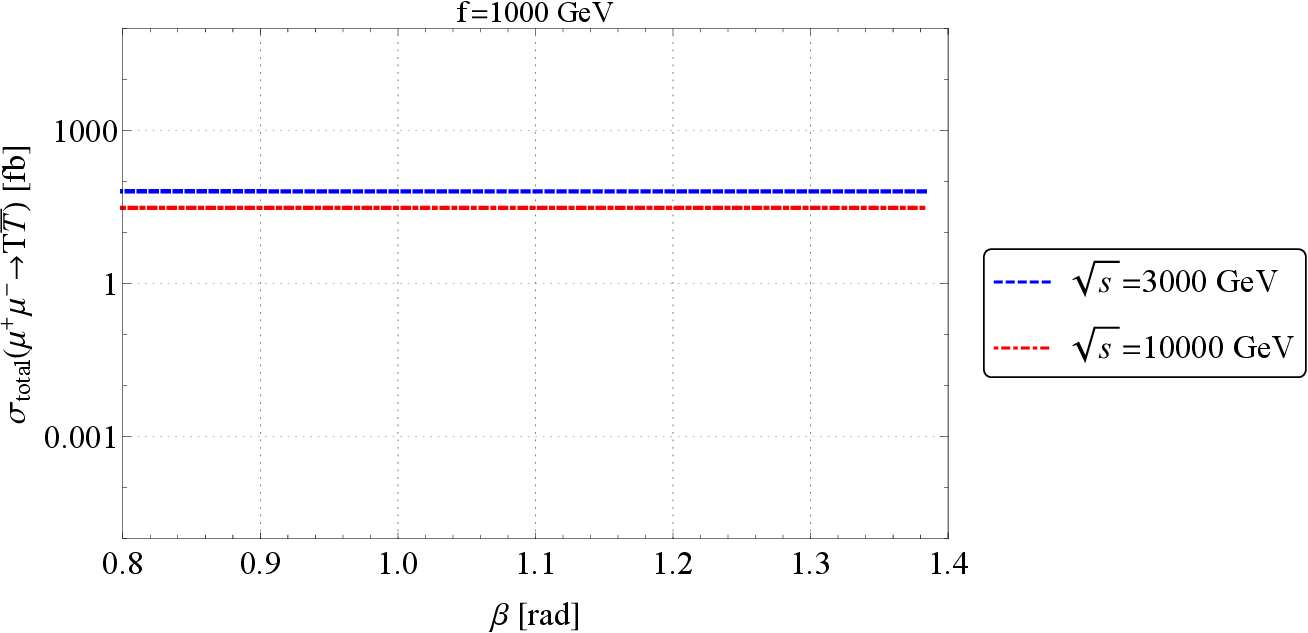}}\\
\subfloat[]{\includegraphics[width=8.0cm]{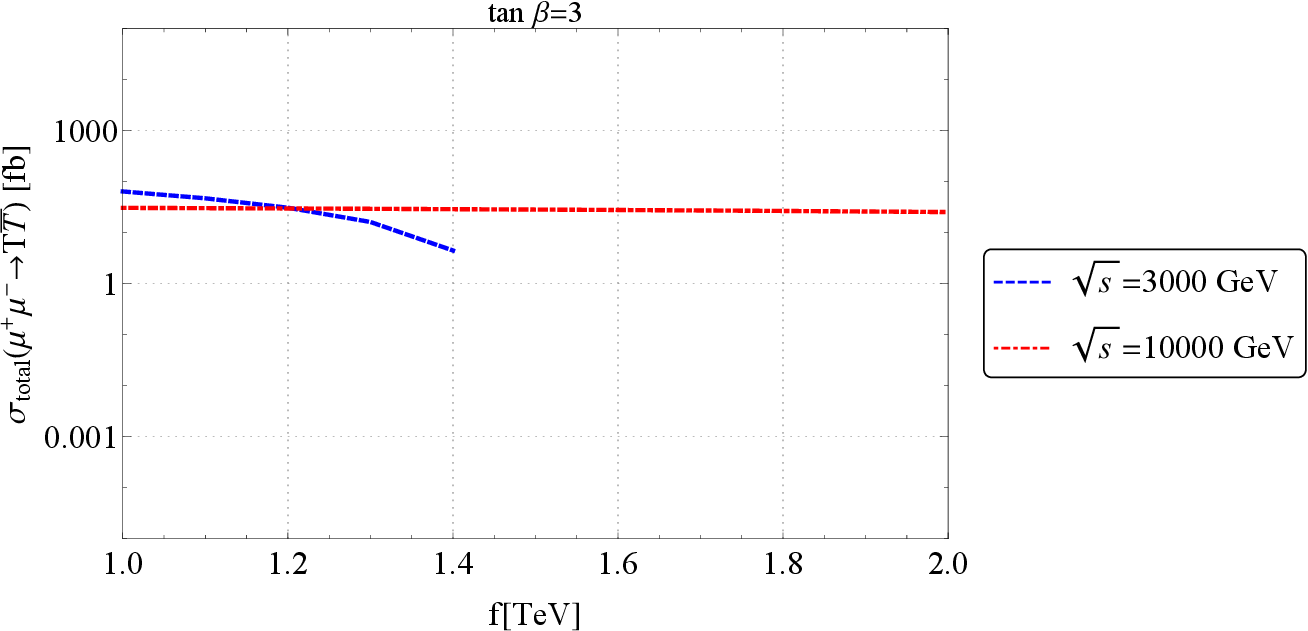}} \vspace{0.6cm}\hspace{0.3cm}
\subfloat[]{\includegraphics[width=8.0cm]{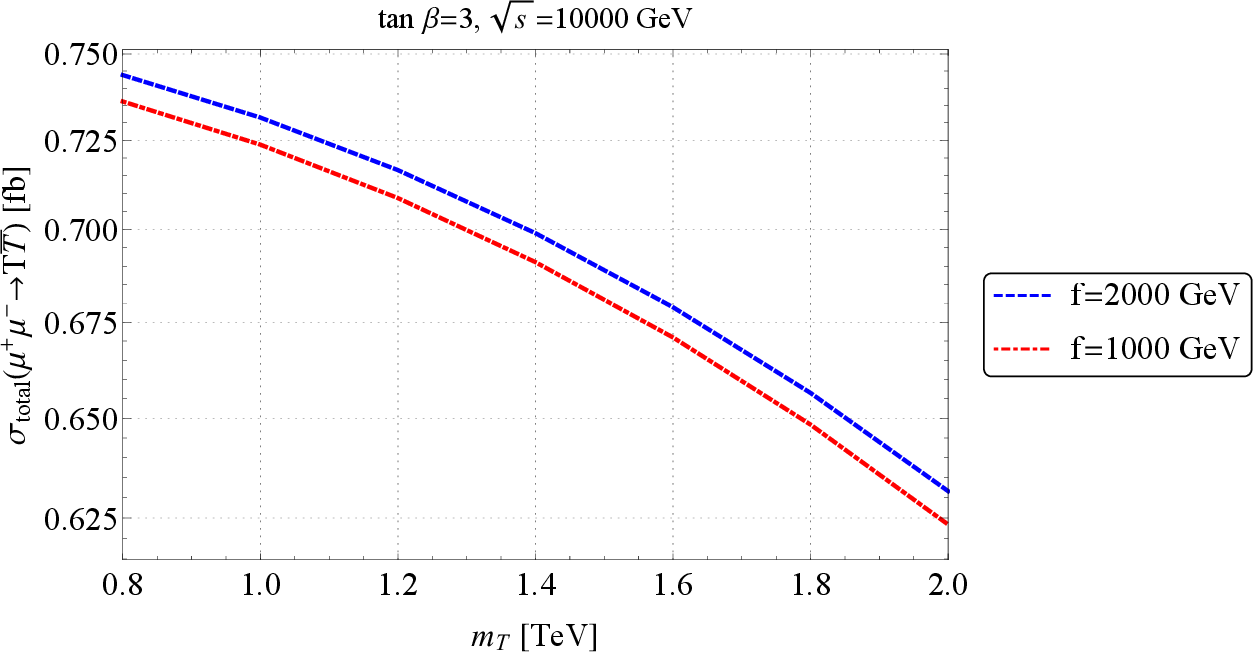}}
\caption{The total cross section of the process $\mu^{+} \mu^{-}  \to T\bar{T}$: (a) as a function of the center-of-mass energy $\sqrt{s}$ (for $f=1000, 2000$ GeV), (b) as a function of the parameter $\beta$ (for $\sqrt{s}=3000, 10000$ GeV), c)  as a function of the energy scale $f$ (for $\sqrt{s}=3000, 10000$ GeV), and d) as a function of $m_{T}$ (for $f=1000, 2000$ GeV).} \label{sigma-tot} 
\end{figure}

We then investigate the impact of varying the free parameters $\sqrt{s}$, $\beta$, $f$, and $m_T$ on the total production cross section of the top partner $T$.  For this purpose, we begin by fixing $\tan\beta = 3$ and analyze  $\sigma_{total}(\mu^{+}\mu^{-} \rightarrow T\bar{T})$ as a function of the center-of-mass energy $\sqrt{s}$ (see~\cref{sigma-tot}(a)). Two distinct curves are generated for  $\sigma_{total}(\mu^{+}\mu^{-} \rightarrow T\bar{T})$, corresponding to the reference values $f = 1000$ GeV and $f = 2000$ GeV, while $\sqrt{s}$ is varied over the range 3000 to  10000 GeV.
For smaller values of the energy scale $f$, larger values of $\sigma_{total}(\mu^{+}\mu^{-} \rightarrow T\bar{T})$  are generated; specifically, when $f=1000$ GeV, the largest peak of the associated curve is reached: $\sigma_{total}(\mu^{+}\mu^{-} \rightarrow T\bar{T})=42399.01$ fb for $\sqrt{s}\approx 4400$ GeV. 
We now examine the effect that the parameters $\beta$ and $\sqrt{s}$ could produce on $\sigma_{total}(\mu^{+}\mu^{-} \rightarrow T\bar{T})$  while fixing the symmetry breaking scale at $f=1000$ GeV. As shown in~\cref{sigma-tot}(b), the pair production cross section of the top partner $T$ is enhanced at lower center-of-mass energies. The two curves shown are generated for $\sqrt{s}= 3000$ GeV and $\sqrt{s} = 10000$ GeV when $\beta \in [\tan^{-1}(1.0), \tan^{-1}(6.0)]$ rad. The largest values of the cross section $\sigma_{total}(\mu^{+}\mu^{-} \rightarrow T\bar{T})$ are obtained for $\sqrt{s}=3000$ GeV:
$\sigma_{total}(\mu^{+}\mu^{-} \rightarrow T\bar{T})=[64.42, 64.04]$ fb. We also observe that there is little dependence of $\sigma_{total}(\mu^{+}\mu^{-} \rightarrow T\bar{T})$ on the value of the parameter $\beta$.
On the other hand, we investigate the behavior of the total cross section  $\sigma_{total}(\mu^{+}\mu^{-} \rightarrow T\bar{T})$ as a function of the new-physics scale $f$, which is varied within the range 1000 to 2000 GeV. As illustrated in~\cref{sigma-tot}(c), the curves have been generated for the center-of-mass energies $\sqrt{s}=3000$ GeV and $\sqrt{s}=10000$ GeV. It can be observed that, for $\sqrt{s}=3000$ GeV, the corresponding curve yields larger cross sections in the interval $f \approx [1000,1200]$ GeV, i.e.,  $\sigma_{total}(\mu^{+}\mu^{-} \rightarrow T\bar{T})=[64.10, 30.78]$ fb. Beyond this region, however, the total cross section gradually decreases, and the curve terminates abruptly at $f \approx 1400$ GeV. This behavior can be understood from kinematic considerations, at $\sqrt{s}=3000$ GeV, the collider no longer provides sufficient center-of-mass energy to produce a pair of top partners $T$, whose masses increase with the symmetry breaking scale $f$ (see~\cref{mT}).
In addition, \cref{sigma-tot}(d) shows the total production cross section  $\sigma_{total}(\mu^{+}\mu^{-} \rightarrow T\bar{T})$ as a function of the top-partner mass $m_{T}$ ($m_{T}\in[800,2000]$ GeV) for two representative values of the symmetry breaking scale, $f=1000$ GeV and $f=2000$ GeV. As can be seen, the total production cross section decreases as the mass of the top partner increases, implying a lower probability for $T\bar{T}$ pair production. This behavior is primarily due to the progressive reduction of the available phase space as $m_{T}$ approaches the kinematic production threshold. Furthermore, the total production cross section exhibits a slight enhancement for larger values of the $f$ scale.
From the figures discussed above, we conclude that the curves associated with $\sigma_{total}(\mu^{+}\mu^{-} \rightarrow T\bar{T})$ decrease as the parameters $\sqrt{s}$, $f$, and $m_{T}$ take on larger values. Furthermore, the cross section $\sigma_{total}(\mu^{+}\mu^{-} \rightarrow T\bar{T})$ exhibits a stronger dependence on the center-of-mass energy $\sqrt{s}$ and $m_{T}$ parameter. In contrast, its dependence on the parameters $f$ and $\beta$ is considerably weaker, as the corresponding values of $\sigma_{total}(\mu^{+}\mu^{-} \rightarrow T\bar{T})$ display only modest variations throughout the parameter ranges considered in our analysis.

From an experimental perspective, and to establish a benchmark for the pair production of the top partner $T$ at a future muon collider, we adopt the design integrated luminosities $\mathcal{L}{_{\mathrm{int}}} = 2000~\mathrm{fb}^{-1}$ and $\mathcal{L}{_{\mathrm{int}}} = 20000~\mathrm{fb}^{-1}$, corresponding to center-of-mass energies of $\sqrt{s}=3000$ GeV and $ \sqrt{s}=10000 $ GeV, respectively. Under these assumptions, \cref{production-TT} presents the number of expected events for $T\bar{T}$ production for the benchmark parameter values $\tan\beta = 3$ and $f = 1000$ GeV.
According to our numerical results, the BLHM contribution to $T\bar{T}$ production at a future muon collider increases with the collider center-of-mass energy. Consequently, the production of the top partner $T$ is a promising channel for probing the BLHM parameter space at future high-energy muon colliders.

\begin{table}[H]
\caption{The total production of $T\bar{T}$ at the muon collider in the context of the BLHM when $\tan\, \beta=3$ and  $f=1 000\ \text{GeV}$.
\label{production-TT}}
    \centering
    \begin{tabular}{|c|c|c|}
    \hline
    \multicolumn{3}{|c|}{$\tan\, \beta=3$, $f=1 000$ GeV} \\
    \hline
         $\mathcal{L}_{\text{int}}$\, [$\text{fb}^{-1}$] & $\sqrt{s}$\,  [GeV] & No. of expected events     \\
         \hline
         2000  & 3000 &  $1.28 \times 10^{5}$ \\
         \hline
         20000 & 10000 & $6.11 \times 10^{5}$   \\
         \hline
    \end{tabular}
\end{table}

\subsection{Single production cross section of the top partner $T$}

\begin{figure}[H]
\center
\subfloat[]{\includegraphics[width=8.0cm]{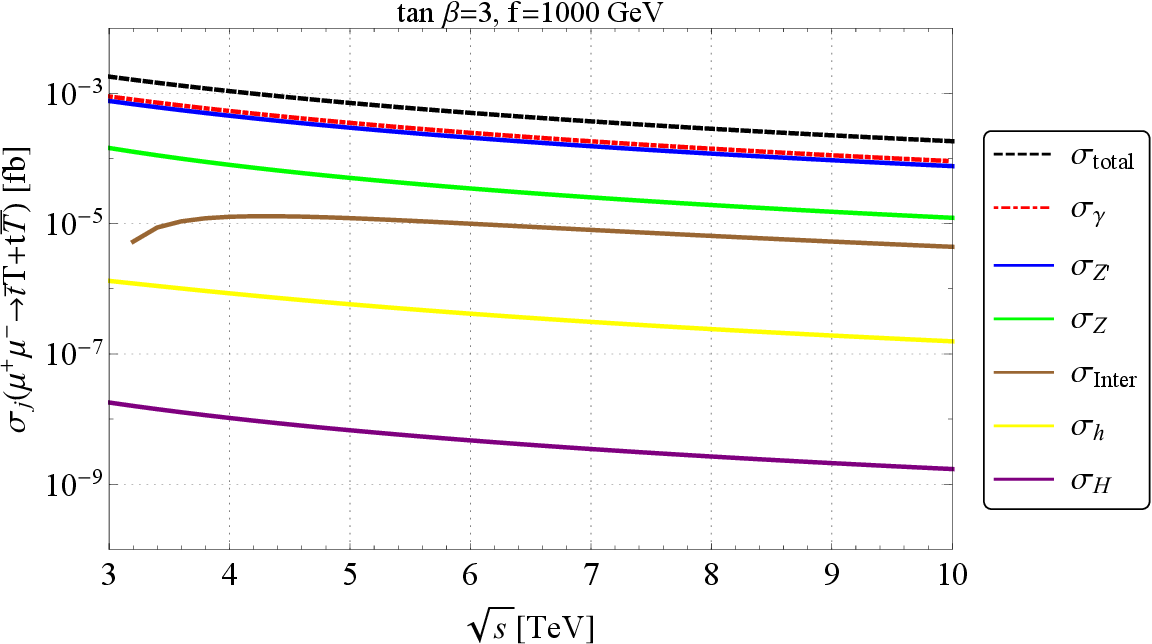}}\vspace{0.6cm}\hspace{0.3cm}
\subfloat[]{\includegraphics[width=8.0cm]{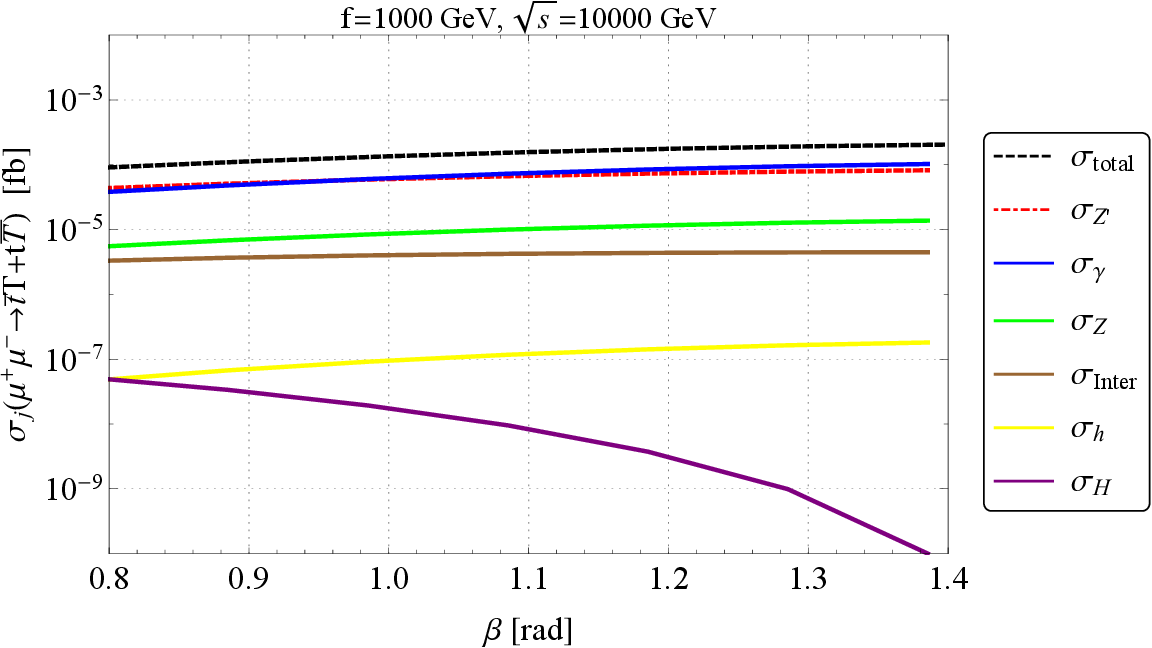}}
\caption{Total and partial cross sections of the process  $\mu^{+} \mu^{-}  \to \bar{t}T+  t\bar{T}$: (a) as a function of the center-of-mass energy $\sqrt{s}$, and (b) as a function of the parameter $\beta$.} \label{sigma-partial-tT} 
\end{figure}

In this subsection, we consider the associated production of a top quark and its heavy partner $T$ through $s$-channel gauge-boson and Higgs-boson exchange. In \cref{sigma-partial-tT}(a), we display the behavior of the individual cross sections  $\sigma_{i}(\mu^{+} \mu^{-} \to  \bar{t}T + t\bar{T})$ together with the total cross section  $\sigma_{total}(\mu^{+} \mu^{-} \to  \bar{t}T + t\bar{T})$ as a function of the collider center-of-mass energy when $\sqrt{s}\in[3000,10000]$ GeV with $\tan \beta=3$ and  $f=1000$ GeV.
We observed that most of the cross sections decrease as the center-of-mass energy $\sqrt{s}$ increases. The interference cross section is the only contribution that exhibits a rising behavior over a particular region of the $\sqrt{s}$ parameter space. In the same figure, it is also evident that the dominant contributions arise from the cross sections $\sigma_{\gamma}(\mu^{+} \mu^{-} \to  \bar{t}T + t\bar{T})$ and $\sigma_{Z'}(\mu^{+} \mu^{-} \to  \bar{t}T + t\bar{T})$, which provide the largest contributions to the total production cross section.
The numerical contributions for these cross sections are $\sigma_{\gamma}(\mu^{+} \mu^{-} \to  \bar{t}T + t\bar{T})=[8.97 \times 10^{-4}, 9.11 \times 10^{-5}]$ fb, $\sigma_{Z'}(\mu^{+} \mu^{-} \to  \bar{t}T + t\bar{T})=[ 7.64 \times 10^{-4}, 7.64 \times 10^{-5}]$ fb, and  $\sigma_{total}(\mu^{+} \mu^{-} \to  \bar{t}T + t\bar{T})=[1.81 \times 10^{-3}, 1.84\times 10^{-4}]$ fb. 
Regarding \cref{sigma-partial-tT}(b), which was generated by fixing $f=1000$ GeV and $\sqrt{s}=10000$ GeV while varying the parameter $\beta$, it can be observed that most of the curves exhibit an increasing behavior throughout the range of $\beta$ considered in our analysis, except for the partial cross section  $\sigma_{H}(\mu^{+} \mu^{-} \to  \bar{t}T + t\bar{T})$.
It is worth noting that the interference contribution  $\sigma_{Inter}(\mu^{+} \mu^{-} \to  \bar{t}T + t\bar{T})$  remains positive in the entire study region of parameter $\beta$, thereby giving rise to constructive interference. 
The figure also shows that the dominant contributions arise from the processes $\mu^{+} \mu^{-} \to Z' \to \bar{t}T + t\bar{T}$ and $\mu^{+} \mu^{-} \to \gamma \to \bar{t}T + t\bar{T}$, whose cross sections are $\sigma_{Z'}(\mu^{+} \mu^{-} \to  \bar{t}T + t\bar{T})=[4.25, 8.20] \times 10^{-4}$ fb and $\sigma_{\gamma}(\mu^{+} \mu^{-} \to  \bar{t}T + t\bar{T})=[3.66 \times 10^{-5}, 1.02 \times 10^{-4}]$ fb, respectively. As for the total cross section, $\sigma_{total}(\mu^{+} \mu^{-} \to  \bar{t}T + t\bar{T})$, its numerical value is $\sigma_{total}(\mu^{+} \mu^{-} \to  \bar{t}T + t\bar{T})=[8.78 \times 10^{-5}, 2.03 \times 10^{-4}]$ fb when $\beta\in[0.8, 1.4]$ rad.
From \cref{sigma-partial-tT}(a) and~\cref{sigma-partial-tT}(b),  it is evident that the contribution of the Higgs boson $H$ gives a small contribution to $\sigma_{total}(\mu^{+} \mu^{-} \to  \bar{t}T + t\bar{T})$ in both benchmark scenarios considered.

\begin{figure}[H]
\center
\includegraphics[width=8.0cm]{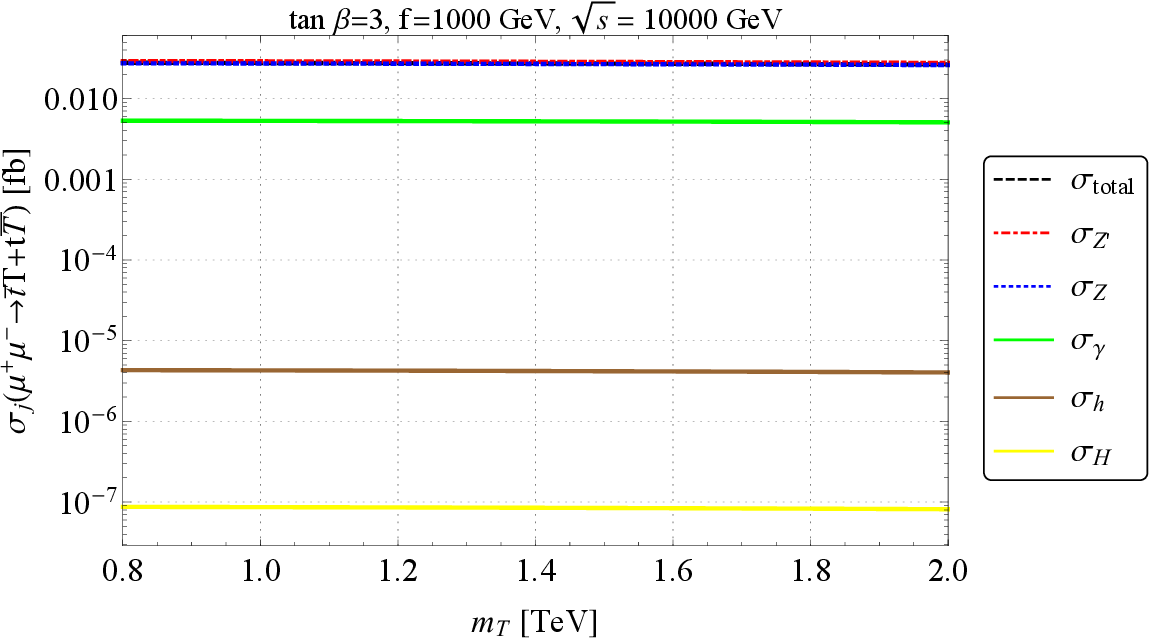}  
\caption{Total and partial cross sections of the process  $\mu^{+} \mu^{-}  \to \bar{t}T+  t\bar{T}$ as a function of $m_{T}$.} \label{sigma-partial-mT-tT} 
\end{figure}

In addition to the parameters $\sqrt{s}$, $f$, and $\beta$, the mass of the top partner $m_{T}$ also plays a significant role in determining the behavior of the production cross section. Accordingly, \cref{sigma-partial-mT-tT} shows the production cross sections $\sigma_{j}(\mu^{+} \mu^{-} \to  \bar{t}T + t\bar{T})$ as functions of $m_{T}$. We find that the $Z'$ mediated channel provides the largest contribution to the production cross section, even exceeding the total cross section $\sigma_{total}(\mu^{+} \mu^{-} \to  \bar{t}T + t\bar{T})$:  $\sigma_{Z'}(\mu^{+} \mu^{-} \to  \bar{t}T + t\bar{T})=[2.91,2.68] \times 10^{-2}$ fb and $\sigma_{total}(\mu^{+} \mu^{-} \to  \bar{t}T + t\bar{T})=[2.77, 2.58] \times 10^{-2}$ fb. This behavior originates from the negative interference term, which remains negative throughout the entire mass range considered and therefore gives rise to destructive interference among the contributing cross sections. The interference contribution has not been displayed in  \cref{sigma-partial-mT-tT}. 
From the corresponding figure, we can also observe that the cross sections of the $\mu^{+} \mu^{-} \to  \bar{t}T + t\bar{T}$ process mediated by gauge bosons provide large numerical contributions compared to those mediated by Higgs bosons.

\begin{figure}[H]
\center
\subfloat[]{\includegraphics[width=8.0cm]{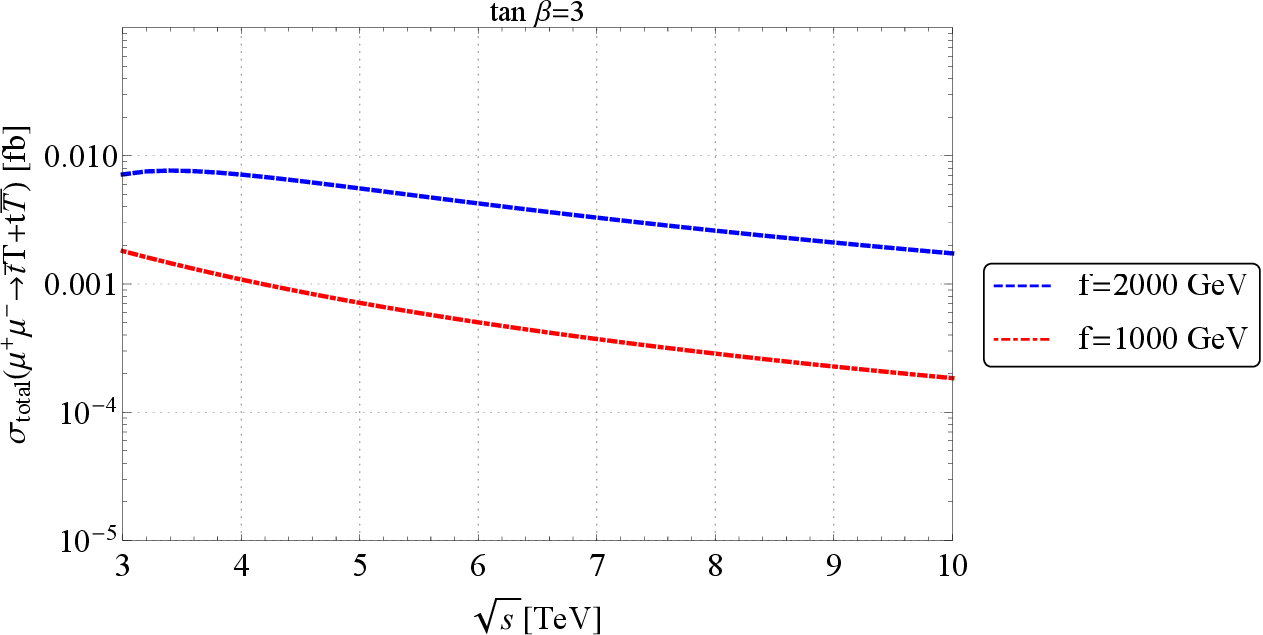}}\vspace{0.6cm}\hspace{0.3cm}
\subfloat[]{\includegraphics[width=8.0cm]{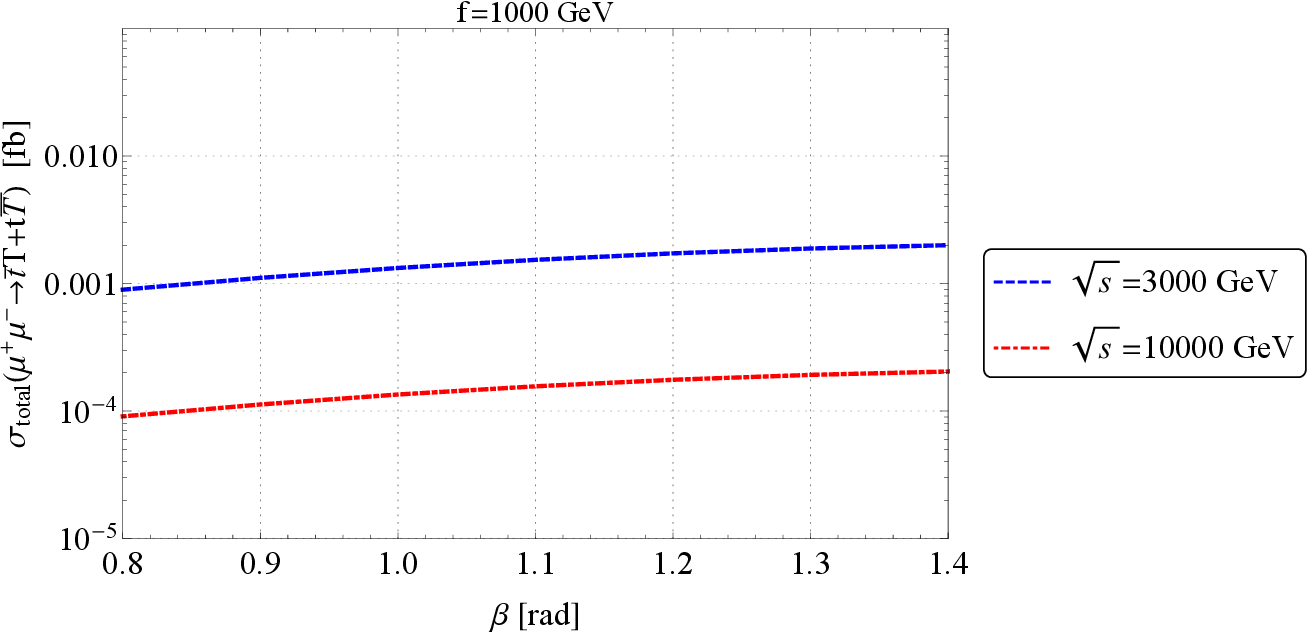}}\\
\subfloat[]{\includegraphics[width=8.0cm]{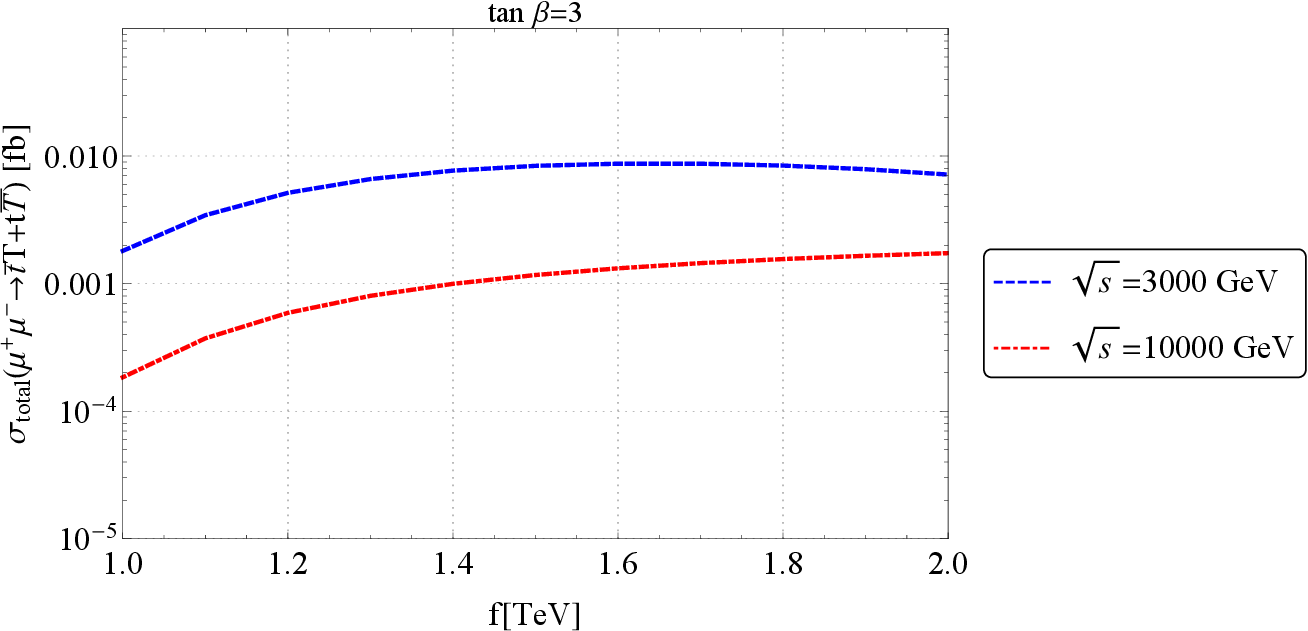}} \vspace{0.6cm}\hspace{0.3cm}
\subfloat[]{\includegraphics[width=8.0cm]{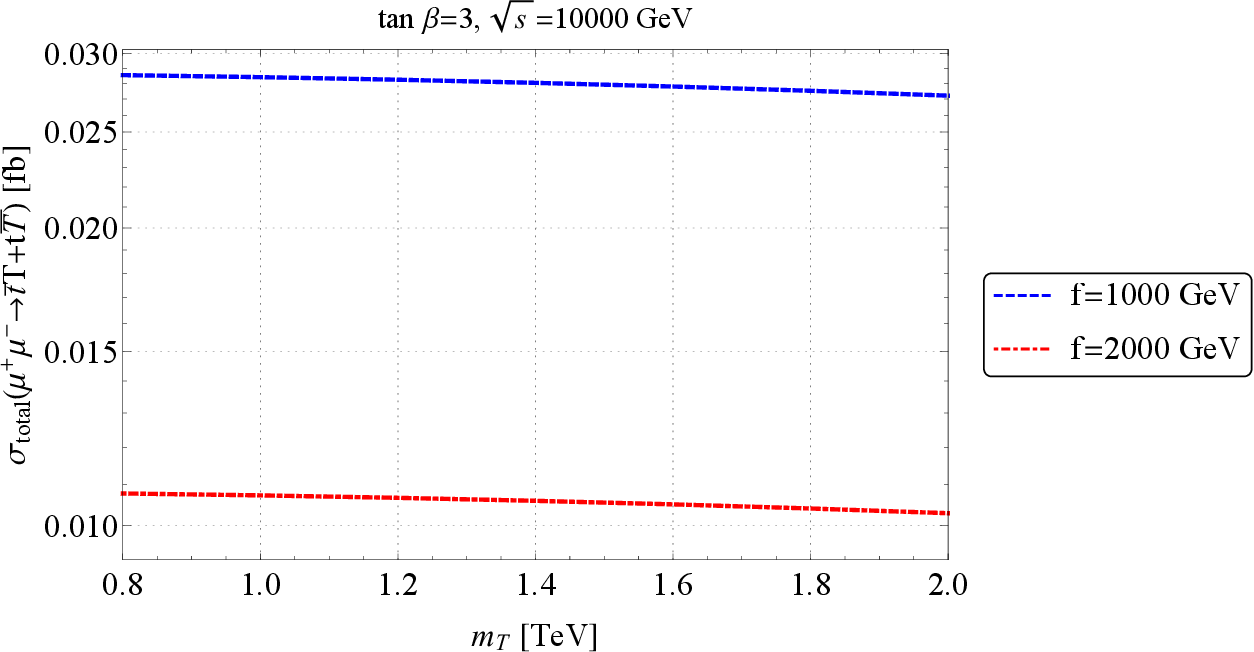}}
\caption{The total cross section of the process  $\mu^{+} \mu^{-}  \to \bar{t}T+  t\bar{T}$: (a) as a function of the center-of-mass energy $\sqrt{s}$ (for $f=1000, 2000$ GeV), (b) as a function of the parameter $\beta$ (for $\sqrt{s}=3000, 10000$ GeV), c)  as a function of the energy scale $f$ (for $\sqrt{s}=3000, 10000$ GeV), and d) as a function of $m_{T}$ (for $f=1000, 2000$ GeV).} \label{sigma-tot-tT} 
\end{figure}

As illustrated in \cref{sigma-tot-tT}, we generate several plots to analyze the dependence of the total cross section $\sigma_{total}(\mu^{+} \mu^{-} \to  \bar{t}T + t\bar{T})$ on the free parameters of the model. We begin by examining the behavior of $\sigma_{total}(\mu^{+} \mu^{-} \to  \bar{t}T + t\bar{T})$ as a function of the center-of-mass energy $\sqrt{s}$ for $\tan\beta = 3$ and two representative values of the $f$ scale, namely $f = 1000$ GeV and $f = 2000$ GeV. As can be seen from \cref{sigma-tot-tT}(a), both curves decrease as the center-of-mass energy approaches $\sqrt{s}=10000$ GeV. More specifically, the total cross section  $\sigma_{total}(\mu^{+} \mu^{-} \to  \bar{t}T + t\bar{T})$ attains larger values for $f=2000$ GeV than for $f=1000$ GeV throughout the energy range considered, i.e.,  $\sigma_{total}(\mu^{+} \mu^{-} \to  \bar{t}T + t\bar{T})=[7.16, 1.73] \times 10^{-3}$ fb while  $\sqrt{s} \in [3000,10000]$ GeV.
We now investigate the impact of the parameters $\beta$ and $\sqrt{s}$ on  $\sigma_{total}(\mu^{+} \mu^{-} \to  \bar{t}T + t\bar{T})$  while fixing the $f$ scale at $f=1000$ GeV. Accordingly, \cref{sigma-tot-tT}(b) displays the curves obtained for the total cross section $\sigma_{total}(\mu^{+} \mu^{-} \to  \bar{t}T + t\bar{T})$, where the center-of-mass energy $\sqrt{s}$  takes specific fixed values while the parameter $\beta$ is varied within the range $0.8 \leq \beta \leq 1.4$ rad.
As shown in the figure, lower center-of-mass energies lead to larger values of $\sigma_{total}(\mu^{+} \mu^{-} \to  \bar{t}T + t\bar{T})$. In particular, for $\sqrt{s}=3000$ GeV, we obtain  $\sigma_{total}(\mu^{+} \mu^{-} \to  \bar{t}T + t\bar{T})=[8.95 \times 10^{-4}, 2.00 \times 10^{-3}]$ fb. 
The two reference curves show an increasing behavior  of up to one order of magnitude as $\beta$ increases up to 1.4 rad.
In \cref{sigma-tot-tT}(c), we study the dependence of  $\sigma_{total}(\mu^{+} \mu^{-} \to  \bar{t}T + t\bar{T})$ on the energy scale $f$, which is varied over the range $1000$  to 2000 GeV while fixing $\tan\beta = 3$. In this case, two curves are shown, corresponding to the center-of-mass energies $\sqrt{s}=3000$ GeV and $\sqrt{s}=10000$ GeV.
As can be seen from the figure, the single production of the top partner $T$ is significantly enhanced at $\sqrt{s}=3000$ GeV, whereas a smaller production rate is obtained at $\sqrt{s}=10000$ GeV: $\sigma_{total}(\mu^{+} \mu^{-} \to  \bar{t}T + t\bar{T})=[1.81,7.16]\times 10^{-3}$ fb and $\sigma_{total}(\mu^{+} \mu^{-} \to  \bar{t}T + t\bar{T})=[1.84 \times 10^{-4}, 1.73 \times 10^{-3}]$ fb, respectively.
Finally, we examine the dependence of the total production cross section $\sigma_{total}(\mu^{+} \mu^{-} \to  \bar{t}T + t\bar{T})$ on the top partner mass $m_{T}$, which is varied over the range 800 to 2000 GeV. For this purpose, two reference curves are generated corresponding to the values $f=1000$ GeV and $f=2000$ GeV. As shown in \cref{sigma-tot-tT}(d), the total production cross section exhibits slightly larger values for the smaller $f$ scale  throughout the entire mass range under consideration.
From the preceding discussion, we conclude that the total cross section $\sigma_{total}(\mu^{+} \mu^{-} \to  \bar{t}T + t\bar{T})$ exhibits a strong dependence on the parameters $\sqrt{s}$, $\beta$, $f$, and $m_{T}$. In particular, $\sigma_{total}(\mu^{+} \mu^{-} \to  \bar{t}T + t\bar{T})$ increases with increasing values of $\beta$ and $f$, whereas it decreases as the center-of-mass energy $\sqrt{s}$ or the mass of the top quark $T$ increases.

As in the previous subsection, \cref{production-tT-f1000,production-tT-f2000} present the expected events for the single production of the top partner $T$ at a future muon collider. The results are obtained for the benchmark parameter sets $\tan\beta = 3$, $f = 1000$ GeV and $\tan\beta = 3$, $f = 2000$ GeV. The events are computed using the design integrated luminosities of the proposed collider, namely $\mathcal{L}{_{\mathrm{int}}} = 2000~\mathrm{fb}^{-1}$ and $\mathcal{L}{_{\mathrm{int}}} = 20000~\mathrm{fb}^{-1}$.
The numerical results indicate that the cross section  $\sigma_{total}(\mu^{+} \mu^{-} \to  \bar{t}T + t\bar{T})$ increases with the collider center-of-mass energy. Nevertheless, the single production of the top partner $T$ is predicted to yield a significantly smaller event rate than the pair production channel, making the latter a more promising avenue for probing the top partner at future high-energy muon colliders.

\begin{table}[H]
\caption{The total production of $ \bar{t}T +  t\bar{T}$ at the muon collider in the context of the BLHM when $\tan\, \beta=3$ and  $f=1 000\ \text{GeV}$.
\label{production-tT-f1000}}
    \centering
    \begin{tabular}{|c|c|c|}
    \hline
    \multicolumn{3}{|c|}{$\tan\, \beta=3$, $f=1 000$ GeV} \\
    \hline
         $\mathcal{L}_{\text{int}}$\, [$\text{fb}^{-1}$] & $\sqrt{s}$\,  [GeV] & No. of expected events     \\
         \hline
         2000  & 3000 &  $ 3 $ \\
         \hline
         20000 & 10000 & $ 3 $   \\
         \hline
    \end{tabular}
\end{table}

\begin{table}[H]
\caption{The total production of $ \bar{t}T +  t\bar{T}$ at the muon collider in the context of the BLHM when $\tan\, \beta=3$ and  $f=2 000\ \text{GeV}$.
\label{production-tT-f2000}}
    \centering
    \begin{tabular}{|c|c|c|}
    \hline
    \multicolumn{3}{|c|}{$\tan\, \beta=3$, $f=2 000$ GeV} \\
    \hline
         $\mathcal{L}_{\text{int}}$\, [$\text{fb}^{-1}$] & $\sqrt{s}$\,  [GeV] & No. of expected events     \\
         \hline
         2000  & 3000 &  $ 14 $ \\
         \hline
         20000 & 10000 & $ 35 $   \\
         \hline
    \end{tabular}
\end{table}

\subsection{The $\mu^{+} \mu^{-} \to T\bar T \to W b W \bar b \to X_W b \bar b$ signal and SM backgrounds}

Thus far, we have carried out a comprehensive phenomenological study of the top partner $T$ within the framework of the BLHM by evaluating its decay widths, branching ratios, and production cross sections for several benchmark configurations of a future muon collider. As a complementary study, in this subsection we perform a signal and background analysis for the pair production of the heavy top partner $T$. In particular, we estimate the signal associated with $T\bar{T}$ production and investigate the relevant SM background processes  to assess the discovery potential of this channel at the collider under consideration.
The signal process considered is  $ \mu^{+} \mu^{-} \to T\bar T \to W b W \bar b \to X_W b \bar b$, where $X_W$ denotes the leptonic $( l^+ \nu_l\, l^- \bar{\nu}_l)$, semi-leptonic $(l^+ \nu_l\, q q')$,  and  hadronic $(q q' q q')$ decay channels. Within the BLHM, this channel represents a subdominant signal of $T\bar{T}$ production. The dominant signal comes from $ \mu^{+} \mu^{-} \to T\bar T \to h W b\,  h W \bar b \to X_W b \bar b b \bar b$, however; this channel exhibits distinctive signatures of new physics.
In the following, we describe the main processes contributing to both the signal of interest and the corresponding SM backgrounds.

{\bf Signal:} \\
  $\hspace{1.5cm}$ $ \mu^{+} \mu^{-} \to T \bar T  \to W b  W \bar b \to \left\{ \begin{array}{lr} l^+ \nu_l\, l^- \bar{\nu}_l b \bar b \\ l^+ \nu_l\, q q' b \bar b \\  q q' q q' b \bar b\end{array} \right.  $, with  $\hspace{0.5cm}$  $l=e, \mu$.   \\

{\bf SM Background Processes:} \\
i) $\mu^{+} \mu^{-} \to t \bar t$,  \\
ii) $\mu^{+} \mu^{-} \to  W W Z$,  \\
iii) $\mu^{+} \mu^{-} \to Z h h$,  \\
iv) $\mu^{+} \mu^{-} \to Z Z h$,  \\
v) $\mu^{+} \mu^{-} \to  W W h$,  \\
vi) $\mu^{+} \mu^{-} \to  h h h$.  \\

\noindent To evaluate the signal cross section, we employ the narrow-width approximation to simplify the calculation. In this context, the total cross section for the signal process, $\sigma\left(  \mu^{+} \mu^{-} \to T\bar T \to W b W \bar b \to X_W b \bar b \right)$, can be expressed as

\begin{eqnarray} \label{N-W}
\sigma\left(  \mu^{+} \mu^{-} \to T\bar T \to W b W \bar b \to X_W b \bar b \right)&\simeq&\sigma\left( \mu^{+} \mu^{-} \to T\bar T \right) \text{Br}(T\to W b )\nonumber \\
&\times& \text{Br} (\bar T \to W \bar b)\, \text{Br}(W) \text{Br}(W),
\end{eqnarray}

\noindent where $\text{Br}(W)$ denotes the branching ratio of the $W$ boson into the decay modes considered in our analysis. We include the leptonic, semileptonic, and hadronic decay channels of the $W$-boson pair when evaluating the signal. \cref{Background-TT} summarizes the representative cross sections of the SM background processes for the signal $ \mu^{+} \mu^{-} \to T\bar T \to W b W \bar b \to X_W b \bar b $  at a future muon collider.
 In this table, we can appreciate that the background processes that contribute significant cross sections are generated by the $\mu^{+} \mu^{-} \to  t\bar{t} $ and $\mu^{+} \mu^{-} \to W W Z$ decay channels.

\begin{table}[H]
\caption{Representative cross sections for the SM background processes of the $\mu^{+} \mu^{-} \to T\bar T \to W b W \bar b \to X_W b \bar b$ signal  at the muon collider for the center-of-mass energies of $\sqrt{s}=3000, 10000\hspace{0.8mm}$ GeV. \label{Background-TT}}
\begin{center}
 \begin{tabular}{|c|c|c|c|}  
\hline\hline
\multicolumn{3}{|c|}{Background cross section [fb]}\\
 \hline\hline
\cline{1-3} {\rm Process}   &  $\sqrt{s}=3000$ GeV  &   $\sqrt{s}=10000$ GeV     \\
\hline
       $\mu^{+} \mu^{-} \to  t\bar{t} $  &    $  5.78 $     &     $ 5.22 \times 10^{-1}  $        \\
\hline
       $\mu^{+} \mu^{-} \to W W Z$  &   $ 2.65  \times 10^{-1}  $      &     $ 2.45 \times 10^{-2}  $        \\
\hline
       $\mu^{+} \mu^{-} \to  Z h  h$  &    $ 2.77 \times 10^{-3}   $     &     $  2.55 \times 10^{-4} $         \\
\hline
       $\mu^{+} \mu^{-} \to ZZ h$    &     $ 3.45 \times 10^{-4} $     &     $ 4.17 \times 10^{-3}  $        \\
\hline
       $\mu^{+} \mu^{-} \to  W W h $  &    $ 3.54 \times 10^{-8} $     &     $ 3.30 \times 10^{-9}  $        \\
\hline
       $\mu^{+} \mu^{-} \to  h h h $  &    $  1.91 \times 10^{-8} $     &     $ 1.82 \times 10^{-9}  $        \\
\hline \hline
\end{tabular}
\end{center}
\end{table}

\noindent \cref{signal-sinFondo-TT,signal-total-TT} provide numerical results for the signal $\mu^{+} \mu^{-} \to T\bar T \to W b  W \bar b \to X_W b\bar b$. In particular,  \cref{signal-sinFondo-TT} lists the signal cross sections without including SM background contributions, whereas \cref{signal-total-TT} presents the corresponding cross sections after accounting for the dominant SM background processes, namely
$\mu^{+} \mu^{-} \to  t\bar t$ and $\mu^{+} \mu^{-} \to WWZ$.  In both scenarios, the reference parameter values have been set to  $f=1000$ GeV and $\tan\, \beta=3$.
From these results, we find that the hadronic decay channel of $T\bar{T}$  yields sizeable production cross sections in both benchmark scenarios considered.
We also observe that the signal cross section for the process $\mu^{+} \mu^{-}  \to T \bar T  \to W b  W \bar b \to X_W b\bar b$ is, in most of the parameter space examined, significantly smaller than the corresponding cross section arising from the SM background processes.
Consequently, the new-physics signal is expected to be substantially suppressed with respect to the SM backgrounds, posing a significant challenge for its experimental observation at a future muon collider.

\begin{table}[H]
\caption{Representative cross sections for the signal  $\mu^{+} \mu^{-}  \to T \bar T  \to W b  W \bar b \to X_W b\bar b$  at a muon collider operating at center-of-mass energies of $\sqrt{s}=3000$, $10000$ GeV. The different decay channels of $W^\pm$ bosons are considered. \label{signal-sinFondo-TT}}
\begin{center}
 \begin{tabular}{|c|c|c|}
\hline\hline
 \multicolumn{2}{|c|}{  $f=1000$ GeV, $\tan\, \beta=3$} \\
\hline 
\multicolumn{2}{|c|}{Signal cross section [fb]: $\sigma(\mu^{+} \mu^{-}  \to T \bar T  \to W b  W \bar b \to q q' q q' b \bar b)$}\\
 \hline
\cline{1-2}  \hspace{1.2cm} $\sqrt{s}=3000$ GeV \hspace{1.2cm}  &   $\sqrt{s}=10000$ GeV     \\
\hline
    $  6.17 \times 10^{-1}  $        &       $  2.94 \times 10^{-1}   $                \\
\hline\hline
\multicolumn{2}{|c|}{Signal cross section [fb]: $\sigma(\mu^{+} \mu^{-}  \to T \bar T  \to W b  W \bar b \to   l^+ \nu_l q q' b\bar  b)$}\\
 \hline
\cline{1-2}  $\sqrt{s}=3000$ GeV  &   $\sqrt{s}=10000$ GeV   \\
\hline
    $  9.93 \times 10^{-2}   $        &       $  4.74 \times 10^{-2} $             \\
\hline\hline
\multicolumn{2}{|c|}{Signal cross section [fb]: $\sigma(\mu^{+} \mu^{-}  \to T \bar T  \to W b  W \bar b \to  l^+ \nu_l\, l^- \bar{\nu}_l b \bar b)$}\\
 \hline
\cline{1-2}  $\sqrt{s}=3000$ GeV  &   $\sqrt{s}=10000$ GeV \\
\hline
    $  1.60 \times 10^{-2}  $        &      $ 7.63 \times 10^{-3}  $            \\
\hline\hline
\end{tabular}
\end{center}
\end{table}

\begin{table}[H]
\caption{Representative cross sections for  the signal  $\mu^{+} \mu^{-}  \to T \bar T  \to W b  W \bar b \to X_W b\bar b$ and dominant backgrounds ($\mu^{+} \mu^{-} \to t\bar{t}/WWZ $) at a muon collider operating at center-of-mass energies of $\sqrt{s}=3000$, $10000$ GeV. The different decay channels of $W^\pm$ bosons are considered. \label{signal-total-TT}}
\begin{center}
 \begin{tabular}{|c|c|c|}
\hline\hline
 \multicolumn{2}{|c|}{  $f=1000$ GeV, $\tan\, \beta=3$} \\
\hline 
\multicolumn{2}{|c|}{Signal cross section [fb]: $\sigma(\mu^{+} \mu^{-}  \to T \bar T  \to W b  W \bar b \to q q' q q' b \bar b)$}\\
 \hline
\cline{1-2}  \hspace{1.2cm} $\sqrt{s}=3000$ GeV \hspace{1.2cm}  &   $\sqrt{s}=10000$ GeV     \\
\hline
    $ 3.04 $        &       $ 5.13 \times 10^{-1}    $                \\
\hline\hline
\multicolumn{2}{|c|}{Signal cross section [fb]: $\sigma(\mu^{+} \mu^{-}  \to T \bar T  \to W b  W \bar b \to   l^+ \nu_l q q' b\bar  b)$}\\
 \hline
\cline{1-2}  $\sqrt{s}=3000$ GeV  &   $\sqrt{s}=10000$ GeV   \\
\hline
    $  4.89 \times 10^{-1}   $        &       $ 8.26 \times 10^{-2}  $             \\
\hline\hline
\multicolumn{2}{|c|}{Signal cross section [fb]: $\sigma(\mu^{+} \mu^{-}  \to T \bar T  \to W b  W \bar b \to  l^+ \nu_l\, l^- \bar{\nu}_l b \bar b)$}\\
 \hline
\cline{1-2}  $\sqrt{s}=3000$ GeV  &   $\sqrt{s}=10000$ GeV \\
\hline
    $ 7.89 \times 10^{-2}   $        &      $ 1.33 \times 10^{-2}  $            \\
\hline\hline
\end{tabular}
\end{center}
\end{table}

\noindent  We also present the number of events for the signal $\mu^{+} \mu^{-}  \to T \bar T  \to W b  W \bar b \to X_W b\bar b$; we take into account the dominant SM background processes:  $ t \bar t$ and $ WWZ$, as shown in \cref{signal-events-TT}.
Our numerical results indicate that the pair production of the BLHM top partner $T$ at a muon collider yields the largest event rates in the hadronic decay channel of $T\bar{T}$.

\begin{table}[H]
\caption{Total production of  $\mu^{+} \mu^{-}  \to T \bar T  \to W b  W \bar b \to X_W b\bar b$ and the background processes  at the muon collider  in the BLHM context.  Leptonic, semi-leptonic, and hadronic  $W^\pm W^\mp$ final states are considered. \label{signal-events-TT}}
\begin{center}
\begin{tabular}{|c|c|c|c|c|}
\hline\hline
 \multicolumn{5}{|c|}{  No. of expected events  } \\
\hline
  & \multicolumn{4}{|c|}{ $f=1000$ GeV, $\tan\, \beta=3$ } \\
\cline{2-5}
 ${\cal L} \, [\rm fb^{-1}]$   &  $\sqrt{s}$ [GeV]   &  Leptonic channel & Semi-leptonic channel  & Hadronic channel \\
\hline\hline
\cline{1-5}
             2000   &    3000     &   158     &     979      &    6079    \\
\hline
          20000    &   10000      &     266   &   1653     &   10 258      \\
\hline
\hline
\end{tabular}
\end{center}
\end{table}

\subsection{The $\mu^{+} \mu^{-} \to  \bar{t}T / t\bar{T} \to W b W \bar b \to X_W b \bar b$ signal and SM backgrounds}

We next investigate the associated production of the top partner $T$ with an SM top quark. For this production mode, the signal process is given by $\mu^{+} \mu^{-}  \to  \bar{t}T / t\bar{T} \to W b  W \bar b \to X_W b\bar b$. Since the top quark decays predominantly through $t \rightarrow W b$, the same SM background processes discussed in the previous subsection contribute to the indicated signal.
Accordingly, \cref{signal-sinFondo-tT} presents representative signal cross sections for the process $\mu^{+} \mu^{-}  \to  \bar{t}T / t\bar{T} \to W b  W \bar b \to X_W b\bar b$ at center-of-mass energies of $\sqrt{s}=3000$ GeV and $\sqrt{s}=10000$ GeV, corresponding to the benchmark operating energies of a future muon collider. \cref{signal-total-tT} also reports the signal cross sections including the dominant SM background contributions (see~\cref{Background-TT}). 
The numerical results presented in both tables were obtained by fixing the free parameters to $f=1000$ GeV and $\tan \beta=3$. Our numerical analysis indicates that the  hadronic and semileptonic decay channels of $\bar{t}T$ $(t\bar{T})$  yield the dominant and subdominant signal cross sections, respectively.
A comparison between \cref{signal-sinFondo-tT,signal-total-tT} further reveals that the SM background cross sections are significantly larger than those predicted for the signal process. Consequently, the new-physics signal is expected to be substantially obscured by the SM backgrounds.

\begin{table}[H]
\caption{Representative cross sections for the signal   $\mu^{+} \mu^{-}  \to  \bar{t}T / t\bar{T} \to W b  W \bar b \to X_W b\bar b$ at a muon collider operating at center-of-mass energies of $\sqrt{s}=3000$, $10000$ GeV. The different decay channels of $W^\pm$ bosons are considered. \label{signal-sinFondo-tT}}
\begin{center}
 \begin{tabular}{|c|c|c|}
\hline\hline
 \multicolumn{2}{|c|}{  $f=1000$ GeV, $\tan\, \beta=3$} \\
\hline 
\multicolumn{2}{|c|}{Signal cross section [fb]: $\sigma(\mu^{+} \mu^{-}  \to  \bar{t}T / t\bar{T}  \to W b  W \bar b \to q q' q q' b \bar b)$}\\
 \hline
\cline{1-2}  \hspace{1.2cm} $\sqrt{s}=3000$ GeV \hspace{1.2cm}  &   $\sqrt{s}=10000$ GeV     \\
\hline
    $ 1.14 \times 10^{-4}  $        &       $  1.17 \times 10^{-5}  $                \\
\hline\hline
\multicolumn{2}{|c|}{Signal cross section [fb]: $\sigma(\mu^{+} \mu^{-}  \to  \bar{t}T / t\bar{T}  \to W b  W \bar b \to   l^+ \nu_l q q' b\bar  b)$}\\
 \hline
\cline{1-2}  $\sqrt{s}=3000$ GeV  &   $\sqrt{s}=10000$ GeV   \\
\hline
    $  1.84 \times 10^{-5}   $        &       $ 1.88 \times 10^{-6} $             \\
\hline\hline
\multicolumn{2}{|c|}{Signal cross section [fb]: $\sigma(\mu^{+} \mu^{-}  \to  \bar{t}T / t\bar{T} \to W b  W \bar b \to  l^+ \nu_l\, l^- \bar{\nu}_l b \bar b)$}\\
 \hline
\cline{1-2}  $\sqrt{s}=3000$ GeV  &   $\sqrt{s}=10000$ GeV \\
\hline
    $  2.97 \times 10^{-6} $        &      $ 3.03 \times 10^{-7}  $            \\
\hline\hline
\end{tabular}
\end{center}
\end{table}

\begin{table}[H]
\caption{Representative cross sections for the signal  $\mu^{+} \mu^{-}  \to  \bar{t}T / t\bar{T} \to W b  W \bar b \to X_W b\bar b$  and dominant backgrounds ($\mu^{+} \mu^{-} \to t\bar{t}/WWZ $)  at a muon collider operating at center-of-mass energies of $\sqrt{s}=3000$, $10000$ GeV. The different decay channels of $W^\pm$ bosons are considered. \label{signal-total-tT}}
\begin{center}
 \begin{tabular}{|c|c|c|}
\hline\hline
 \multicolumn{2}{|c|}{  $f=1000$ GeV, $\tan\, \beta=3$} \\
\hline 
\multicolumn{2}{|c|}{Signal cross section [fb]: $\sigma(\mu^{+} \mu^{-}  \to  \bar{t}T / t\bar{T}  \to W b  W \bar b \to q q' q q' b \bar b)$}\\
 \hline
\cline{1-2}  \hspace{1.2cm} $\sqrt{s}=3000$ GeV \hspace{1.2cm}  &   $\sqrt{s}=10000$ GeV     \\
\hline
    $ 2.42  $        &       $ 2.19 \times 10^{-1}   $                \\
\hline\hline
\multicolumn{2}{|c|}{Signal cross section [fb]: $\sigma(\mu^{+} \mu^{-}  \to  \bar{t}T / t\bar{T}  \to W b  W \bar b \to   l^+ \nu_l q q' b\bar  b)$}\\
 \hline
\cline{1-2}  $\sqrt{s}=3000$ GeV  &   $\sqrt{s}=10000$ GeV   \\
\hline
    $   3.90 \times 10^{-1}  $        &       $ 3.52 \times 10^{-2} $             \\
\hline\hline
\multicolumn{2}{|c|}{Signal cross section [fb]: $\sigma(\mu^{+} \mu^{-}  \to  \bar{t}T / t\bar{T} \to W b  W \bar b \to  l^+ \nu_l\, l^- \bar{\nu}_l b \bar b)$}\\
 \hline
\cline{1-2}  $\sqrt{s}=3000$ GeV  &   $\sqrt{s}=10000$ GeV \\
\hline
    $ 6.29 \times 10^{-2}  $        &      $  5.68 \times 10^{-3} $            \\
\hline\hline
\end{tabular}
\end{center}
\end{table}

\cref{signal-events-tT} presents the expected event yields for the signal process  $\mu^{+} \mu^{-}  \to \bar{t}T / t\bar{T}  \to W b  W \bar b \to X_W b\bar b$ after including the dominant SM background processes.
Our  results indicate that the single production of the BLHM top partner $T$ at the muon collider is expected to yield the largest event rates in the fully hadronic decay channel of $ \bar{t}T / t\bar{T}$.

\begin{table}[H]
\caption{Total production of  $\mu^{+} \mu^{-}  \to \bar{t}T / t\bar{T}  \to W b  W \bar b \to X_W b\bar b$ and the background processes  at the muon collider  in the BLHM context.  Leptonic, semi-leptonic, and hadronic  $W^\pm W^\mp$ final states are considered. \label{signal-events-tT}}
\begin{center}
\begin{tabular}{|c|c|c|c|c|}
\hline\hline
 \multicolumn{5}{|c|}{  No. of expected events  } \\
\hline
  & \multicolumn{4}{|c|}{ $f=1000$ GeV, $\tan\, \beta=3$ } \\
\cline{2-5}
 ${\cal L} \, [\rm fb^{-1}]$   &  $\sqrt{s}$ [GeV]   &  Leptonic channel & Semi-leptonic channel  & Hadronic channel \\
\hline\hline
\cline{1-5}
             2000   &    3000     &   126      &   781     &   4846    \\
\hline
          20000    &   10000      &    114      &   705     &    4376    \\
\hline
\hline
\end{tabular}
\end{center}
\end{table}

\section{Conclusions} \label{sec:conclusions}

In this work, we have investigated the single and pair production of the top partner $T$ at a future muon collider within the framework of the BLHM. This extension of the SM provides an appealing solution to unresolved issues such as the hierarchy problem. Thus, the BLHM naturally predicts the existence of new particles whose contributions cancel the quadratic divergences in the Higgs boson mass, thereby stabilizing the electroweak scale and preserving a naturally light Higgs boson.

Our study of the heavy quark $T$ predicted by the BLHM focused on the impact of the model parameters $f$, $\beta$, and $m_{T}$ on its decay widths and branching ratios. We also investigated the production cross sections for the processes $\mu^{+}\mu^{-} \to (\gamma, Z, Z', h, H) \rightarrow  T\bar{T}$ and $\mu^{+}\mu^{-} \to (\gamma, Z, Z', h, H) \rightarrow \bar{t}T + t\bar{T}$ at the benchmark center-of-mass energies and integrated luminosities of the muon collider. Furthermore, we examined the sensitivity of these production cross sections to variations in the parameters $f$, $\beta$, $\sqrt{s}$, and $m_T$ over the parameter ranges considered in this paper.  Our numerical results indicate that the pair production of the top partner $T$ constitutes  the most promising discovery channel for probing physics beyond the SM and exploring the parameter space of the BLHM at a future high-energy muon collider. 
In contrast, the associated production channel, $\mu^{+}\mu^{-}\rightarrow \bar{t}T+t\bar{T}$, yields considerably smaller production cross sections and is therefore expected to provide a less favorable channel for the experimental search for the top partner $T$.

We also performed a phenomenological analysis of the characteristic signal associated with the single and pair production mechanisms of the top partner $T$, providing approximate estimates of the corresponding signal cross sections for the processes 
$\mu^{+} \mu^{-}  \to \bar{t}T / t\bar{T}  \to W b  W \bar b \to X_W b\bar b$ 
and  $\mu^{+} \mu^{-}  \to T\bar{T}  \to W b  W \bar b \to X_W b\bar b$, respectively.
Our numerical results indicate that pair production of the heavy quark $T$ at a future muon collider provides a more favorable channel for probing this particle than its associated production with a SM top quark. Although the signal cross section for the pair production channel is smaller than the corresponding SM background cross section over certain regions of the parameter space, the difference remains moderate compared with the associated production channel. Consequently, the new-physics signal is expected to be partially obscured by the SM backgrounds, making its experimental observation challenging but still potentially feasible with an optimized event selection strategy. In contrast, for the associated production process, the SM background cross sections exceed the signal cross sections by a substantial margin throughout the parameter space considered, implying that the signal is expected to be largely overwhelmed by the SM backgrounds.

\cref{signal-events-TT} and~\cref{signal-events-tT} also present the expected events for the signal processes   $\mu^{+} \mu^{-}  \to T\bar{T}  \to W b  W \bar b \to X_W b\bar b$ and $\mu^{+} \mu^{-}  \to \bar{t}T / t\bar{T}  \to W b  W \bar b \to X_W b\bar b$ 
including the dominant SM background contributions. In both production modes, the hadronic decay channel yields the largest event rates for the benchmark center-of-mass energies considered, making it the most promising final state for future experimental searches at a future muon collider.

\vspace{2.0cm}

\begin{center}
{\bf ACKNOWLEDGMENTS}
\end{center}

 E. Cruz-Albaro appreciates the postdoctoral stay at the
Universidad Autónoma de Zacatecas. A.G.R. and  D.E.G. thank SNII and PROFEXCE (M\'exico).

\vspace{1cm}

\begin{center}
   {\bf DATA AVAILABILITY}
\end{center}

  All data generated or analyzed during this study are included in this article.


\newpage

\appendix

\section{Transition amplitudes for the processes $\mu^{+} \mu^{-} \to  \bar{T} T $ and $\mu^{+} \mu^{-} \to  t \bar{T}\, (\bar{t} T) $ } \label{app:amplitudes}

In this Appendix, we provide the different transition amplitudes for the pair and single production of the heavy  quark $T$ (see~\cref{fig:TT}). These production mechanisms of the $T$ quark are generated through the processes $\mu^+\mu^-\to (\gamma, Z, Z', h, H ) \to \bar{T}T$ and $\mu^+\mu^-\to (\gamma, Z, Z', h, H ) \to t \bar{T}$, for which the corresponding amplitudes are given in \cref{eq:MTTi,eq:MTTi1,eq:MTTi2,eq:MTTi3,eq:MTTf} and~\cref{MTti,MTti1,MTti2,MTti3,MTtf}, respectively.
Concerning the other mechanism of single production of the $T$ quark via the processes $\mu^+\mu^-\to (\gamma, Z, Z', h, H ) \to \bar{t}T$. As mentioned above, the corresponding contributions are  identical to those obtained for the processes $\mu^+\mu^-\to (\gamma, Z, Z', h, H ) \to t\bar{T}$. Consequently, the transition amplitudes for the $\mu^+\mu^-\to (\gamma, Z, Z', h, H ) \to \bar{t}T$ processes can be derived from \cref{MTti,MTti1,MTti2,MTti3,MTtf} by replacing $t\to T$ and $\bar{T}\to \bar{t}$.

\begin{align} \label{eq:MTTi}
    \mathcal{M}_{\gamma}(\mu^+\mu^-\to T\bar{T}) =& \left(\frac{g_{\alpha\beta}}{q^2}\right)\bar{U}_{r_4}(k_2)\gamma^{\beta}\left(g_V^{T\bar{T}\gamma}+g_A^{T\bar{T}\gamma}\gamma^5\right)V_{r_3}(k_1)\nonumber\\
    &\times  \bar{V}_{r_2}(p_2)\gamma^{\alpha} \left(g_V^{\mu\mu\gamma}+g_A^{\mu\mu\gamma}\gamma^5\right) U_{r_1}(p_1),
\end{align}

\begin{align} \label{eq:MTTi1}
	\mathcal{M}_Z(\mu^+\mu^-\rightarrow T\bar{T})=& \left(\frac{1}{q^2-m_Z^2+i\Gamma_Zm_Z}\left(g_{\alpha\beta}-\frac{q_{\alpha}q_{\beta}}{m_Z^2}\right)\right)\bar{U}_{r_4}(k_2)\gamma^{\beta} \left(g_V^{T\bar{T}Z}+g_A^{T\bar{T}Z}\gamma^5\right) \nonumber\\
	&\times V_{r_3}(k_1) \bar{V}_{r_2}(p_2)\gamma^{\alpha}\left(g_V^{Z \mu\mu }+g_A^{Z \mu\mu }\gamma^5\right)U_{r_1}(p_1),
\end{align}

\begin{align}  \label{eq:MTTi2}
    \mathcal{M}_{Z'}(\mu^+\mu^-\rightarrow T\bar{T})=&  \left(\frac{1}{q^2-m_{Z'}^2+i\Gamma_{Z'}m_{Z'}}\left(g_{\alpha\beta}-\frac{q_{\alpha}q_{\beta}}{m_{Z'}^2}\right)\right)\bar{U}_{r_4}(k_2)\gamma^{\beta} \left(g_V^{T\bar{T}Z'}+g_A^{T\bar{T}Z'}\gamma^5\right) \nonumber\\
	&\times V_{r_3}(k_1)   \bar{V}_{r_2}(p_2)\gamma^{\alpha
	}\left(g_V^{Z' \mu\mu }+g_A^{Z' \mu\mu }\right)U_{r_1}(p_1),
\end{align}

\begin{equation}  \label{eq:MTTi3}
    \mathcal{M}_h(\mu^+\mu^-\to T\bar{T})= \frac{g_{hTT}g_{h\mu\mu}}{s-m_h^2+im_h\Gamma_h}\bar{U}(p_2)V(p_1)\bar{V}(k_2)U(k_1),
\end{equation}

\begin{align} \label{eq:MTTf}
    \mathcal{M}_H(\mu^+\mu^-\to T\bar{T}) =& -\frac{g_{HTT}g_{H\mu\mu}}{s-m_H^2+im_H\Gamma_H}\bar{U}(p_2)V(p_1)  \bar{V}(k_2)U(k_1),
\end{align}

\begin{align}\label{MTti}
   \mathcal{M}_{\gamma}(\mu^+\mu^-\to t\bar{T}  )=&- \bigg(\frac{g_{\alpha\beta}}{q^2}\bigg)\bar{U}_{r_4}(k_2)\gamma^{\beta}\left(g_V^{\bar{T}t\gamma}+g_A^{\bar{T}t\gamma}\gamma^5\right)V_{r_3}(k_1)
	\bar{V}_{r_2}(p_2)\gamma^{\alpha}\left(g_V^{\mu\mu\gamma}+g_A^{\mu\mu\gamma}\gamma^5\right) \nonumber\\
    &\times U_{r_1}(p_1),
\end{align}

\begin{align} \label{MTti1}
    \mathcal{M}_Z(\mu^+\mu^-\to t\bar{T})=&\left(\frac{1}{q^2-m_Z^2+im_Z\Gamma_Z}\left(g_{\alpha\beta}-\frac{q_{\alpha}q_{\beta}}{m_Z^2}\right)\right)\bar{U}_{r_4}(k_2)\gamma^{\beta}\left(g_V^{t \bar{T} Z}+g_A^{t \bar{T} Z}\gamma^5\right)V_{r_3}(k_1)\nonumber\\
    &\times \bar{V}_{r_2}(p_2)\gamma^{\alpha}\left(g_V^{Z \mu\mu }+g_A^{Z \mu\mu }\gamma^5\right)U_{r_1}(p_1),
\end{align}

\begin{align} \label{MTti2}
    \mathcal{M}_{Z'}(\mu^+\mu^-\to t\bar{T})=& \left(\frac{1}{q^2-m_{Z'}^2+im_{Z'}\Gamma_{Z'}}\left(g_{\alpha\beta}-\frac{q_{\alpha}q_{\beta}}{m_{Z'}^2}\right)\right)\bar{U}_{r_4}(k_2)\gamma^{\beta}\left(g_V^{t \bar{T} Z'}+g_A^{t\bar{T}Z'}\gamma^5\right) \nonumber\\
	&\times V_{r_3}(k_1) \bar{V}_{r_2}(p_2)\gamma^{\alpha}\left(g_V^{Z' \mu\mu }+g_A^{Z' \mu\mu}\gamma^5\right)U_{r_1}(p_1),
\end{align}

\begin{align} \label{MTti3}
    \mathcal{M}_h(\mu^+\mu^-\to t\bar{T}) =& \frac{g_{h\mu\mu}}{s-m_h^2+im_h\Gamma_h}\bar{U}(p_2)(g_V^{t \bar{T} h}+g_A^{t \bar{T} h}\gamma^5)V(p_1) \bar{V}(k_2)U(k_1),
\end{align}
 
\begin{align} \label{MTtf}
    \mathcal{M}_H(\mu^+\mu^-\to t\bar{T}) =& -\frac{g_{H\mu\mu}}{s-m_H^2+im_H\Gamma_H}\bar{U}(p_2)(g_V^{t \bar{T} H}+g_A^{t \bar{T} H}\gamma^5)V(p_1) \nonumber\\
    & \times  \bar{V}(k_2)U(k_1).
\end{align}

\section{Feynman's Rules} \label{app:rulesF}

The effective couplings, as well as the vector and axial-vector couplings involved in our calculations, are presented explicitly in this Appendix.
For convenience, we also adopt the following shorthand notation:

\begin{eqnarray}
c_{\alpha} &=& \cos \alpha, \\
s_{\alpha} &=& \sin \alpha, \\
c_{\beta} &=& \cos \beta, \\
 s_{\beta} &=& \sin \beta, \\
 y_{\mu}&=& \frac{m_\mu}{v\,  s_\beta} \left(1-\frac{v^{2}}{3 f^{2}} \right)^{-1/2}.
\end{eqnarray}


\begin{table}[H]
\caption{Vector couplings derived from interaction vertices involving two fermions and either a gauge boson or a Higgs boson.}
\label{Table-ii}
\begin{tabular}{|p{1.8cm}| p{14.0cm}| p{7.5cm}|}
\hline
\hline
  \textbf{Vertex}    &   \hspace{5cm}\textbf{Vector couplings}\\
\hline
\hline

$t \bar{T}\gamma$ & $g^{t \bar{T}\gamma}_{V}=\frac{gs_W}{f^2(y_1^2+y_2^2)^2(y_1^2+y_3^2)^{3/2}(-y_2^2+y_3^2)}[2y_1y_2(y_1^2+y_2^2)\sqrt{y_1^2+y_3^2}(-(f^2(y_2^2-y_3^2)(y_1^2+y_3^2))+2v^2(2y_2^2y_3^2-5y_3^4+y_1^2(8y_2^2-2y_3^2)))-12v^2y_1y_2(y_1^2y_2^2)y_3^2\sqrt{y_1^2+y_3^2}(2y_1^2-2y_2^2+y_3^2)c_{2\beta}-fv(2y_1^2-y_2^2)y_3(y_2^2-y_3^2)(8y_1^4+8y_2^2y_3^2-3\sqrt{y_1^2+y_3^2}+8y_1^2(y_2^2+y_3^2))s_{\beta}]$ \\
        \hline
$t \bar T Z$ & $g^{t \bar T  Z}_{V}=\displaystyle \frac{g}{24c_Wf^2(y_1^2+y_2^2)^2(y_1^2+y_3^2)^{3/2}(y_3^2-y_2^2)}
[8s_W^2v^2y_1y_2(y_1^2+y_2^2)\sqrt{y_1^2+y_3^2}(4c_{\beta}^2(y_1^2+y_3^2)(y_2^2-y_3^2)+s_{\beta}^2(2y_1^2-y_3^2)(2y_2^2+y_3^2))-fs_{\beta}vy_3(2y_1^2-y_2^2)(y_2^2-y_3^2)(3c_W^2\sqrt{y_1^2+y_3^2}+8s_W^2(y_1^2+y_2^2)(y_1^2+y_3^2))+2f^2s_W^2y_1y_2(y_1^2+y_2^2)(y_1^2+y_3^2)^{3/2}(y_3^2-y_2^2)]$\\
        \hline
$t \bar T Z'$ & $g^{t \bar T  Z'}_{V}=\frac{gs_\beta v(2y_1^2-y_2^2)y_3}{8f(y_1^2+y_2^2)^2(y_1^2+y_3^2)} $\\
        \hline
$t \bar{T} h$ & $g^{t \bar{T} h}_{V}=\frac{c_{\alpha}}{2f(y_1^2+y_2^2)^{3/2}(y_1^2+y_3^2)^{3/2}}[-(f(2y_1^4+y_1^2y_2^2-y_2^4)y_3(y_1^2+y_3^2))+vy_1\sqrt{y_1^2+y_3^2}(2y_1^4(y_2-y_3)+2y_2^2y_3(y_2-y_3)^2+y_1^2(2y_2^3+5y_2y_3^2+2y_3^3))s_{\beta}]$ \\
        \hline
		$t \bar T H$ &   $g_V^{t \bar T H}=\frac{1}{12f^2(y_1^2+y_2^2{5/2}(y_2^2-y_3^2)(y_1^2+y_3^2)^{3/2}}[6f^2s_{\alpha}(2y_1^2-y_2^2)(y_1^2+y_2^2)^2y_3(y_2^2-y_3^2)(y_1^2+y_3^2)-2s_{\alpha}v^2(2y_1^2-y_2^2)(y_1^2+y_2^2)^2y_3(y_2^2-y_3^2)(y_1^2-y_3^2)(y_1^2+y_3^2)c_{\beta}^2-6fs_{\alpha}vy_1(y_1^2+y_2^2)(y_2^2-y_3^2)\sqrt{y_1^2+y_3^2}(2y_1^4(y_2-y_3)+2y_2^2(y_2-y_3)^2y_3+y_1^2(2y_2^3+5y_2^2y_3^2+2y_3^3))s_{\beta}-c_{\alpha}v^2y_3(-20y_1^8(y_2^2-y_3^2)+3y_2^4\sqrt{y_1^2+y_3^2}(-y_2^2+y_3^2)+4y_1^6(-6y_2^4+y_2^2y_3^2+5y_3^4)+y_1^2(-4y_2^6y_3^2+9y_2^4\sqrt{y_1^2+y_3^2}-9y_2^2y_3^2\sqrt{y_1^2+y_3^2})+6y_1^4(-6y_2^4y_3^2+y_3^2\sqrt{y_1^2+y_3^2}-y_2^2(-4y_3^4+\sqrt{y_1^2+y_3^2})))c_\beta
		s_{\beta}-3s_\alpha^2y_3(4y_1^8(2y_2^+y_3^2)+2y_1^6(5y_2^4+11y_2^2y_3^2+2y_3^4)+y_1^2y_2^4(-6y_2^4+2y_2^2y_3^2+16y_3^4-3\sqrt{y_1^2+y_3^2})-y_2^4(-y_2^2+y_3^2)(-6y_2^2y_3^2+\sqrt{y_1^2+y_3^2})+2y_1^4(-2y_2^6+13y_2^4y_3^2+2y_3^8\sqrt{y_1^2+y_3^2}+y_2^2(7y_3^4+\sqrt{y_1^2+y_3^2})))s_{\beta}^2-2c_{\alpha}v^2y_2^4y_3(3y_1^4y_2^2+4y_2^2y_3^2(y_1^2-y_3^2)+y_1^2(4y_2^4-3y_3^4))s_{2\beta}]$\\
		\hline
		$\overline{T}T\gamma$    &   $g_V^{\overline{T}T\gamma}=\frac{gs_W}{6f^2(y_1^2+y_2^2)^2(y_1^2+y_3^2)(y_2^2-y_3^2)}[4f^2(y_1^2+y_2^2)^2(y_1^2+y_3^2)(y_1^2-y_3^2)-v^2y_1^2c_{2\beta}(y_1^4(y_2^2(12y_2^2-1)-3y_3^4+y_3^2)+y_1^2(2y_2^4(6y_3^2-1)+y_2^2y_3^2(9y_2^2+5)-3(y_3^6+y_3^4))-y_2^2(y_2^4+y_2^2(5y_3^2-12y_3^4)+3y_3^4(y_3^2-2)))+v^2(y_1^6(8y_2^4-4y_2^2y_3^2+y_2^2+5y_3^4-y_3^2)+y_1^4(8y_2^6+y_2^4(4y_3^2+2)+y_2^2y_3^2(y_3^2+3)+5y_3^4(y_3^2-1))+y_1^2(y_2^6(8y_3^2+1)-y_2^4y_3^2(4y_3^2+3)+y_2^2y_3^4(5y_3^2+2))+2y_2^4y_3^2(y_1^2-y_3^2)]$\\
		\hline
	$T \bar{T}Z $  &   $g_V^{T \bar{T}Z }=  \frac{g}{6c_W}[3c_W^2-s_W^2-\frac{4s_Wv^2y_1^2(s_{2\beta}^2(2y_2^2+y_3^2)^2+4c_\beta^2(y_2^2-y_3^2)^2)}{f^2(y_1^2+y_2^2)(y_2^2-y_3^2)}] $\\
	\hline
	$T \bar{T}Z' $  &   $g_V^{T \bar{T}Z' }= 0 $\\
	\hline
$Z \mu^{+}\mu^{-} $  &   $g_V^{Z \mu \mu }= \frac{g}{c_W}(-\frac{1}{4}+s_W^2) $\\
	\hline
	$Z' \mu^{+}\mu^{-} $  &   $g_V^{Z' \mu \mu }= \frac{g}{4} $\\
	\hline
$\gamma \mu^{+}\mu^{-} $  &   $g_V^{\mu \mu\gamma}=e$\\
		\hline
\end{tabular}
\end{table}

\begin{table}[H]
\caption{Axial-vector couplings derived from interaction vertices involving two fermions and either a gauge boson or a Higgs boson.}
\label{Table-iii}
\begin{tabular}{|p{1.8cm}| p{14.0cm}| p{7.5cm}|}
\hline
\hline
  \textbf{Vertex}    &    \hspace{5cm} \textbf{Axial-vector couplings}\\
\hline
\hline

$T\overline{t}\gamma$ & $g^{T\overline{t}\gamma}_{A}=-\frac{gs_W}{24f^2(y_1^2+y_2^2)^2(y_1^2+y_3^2)^{3/2}(-y_2^2+y_3^2)}[2y_1y_2(y_1^2+y_2^2)\sqrt{y_1^2+y_3^2}(-(f^2(y_2^2-y_3^2)(y_1^2+y_3^2))+2v^2(2y_2^2y_3^2-5y_3^4+y_1^2(8y_2^2-2y_3^2)))-12v^2y_1y_2(y_1^+y_2^2)y_3^2\sqrt{y_1^2+y_3^2}(2y_1^2-2y_2^2+y_3^2)c_{2\beta}-fv(2y_1^2-y_2^2)y_3(y_2^2-y_3^2)(8y_1^4+8y_2^2y_3^2-3\sqrt{y_1^2+y_3^2}+8y_1^2(y_2^2+y_3^2))s_{\beta}]$ \\
        \hline
$t \bar T Z$ &   $g^{t \bar T Z}_{A}=\displaystyle \frac{g}{24c_Wf^2(y_1^2+y_2^2)^2(y_1^2+y_3^2)^{3/2}(y_3^2-y_2^2)}[-8s_W^2v^2y_1y_2(y_1^2+y_2^2)\sqrt{y_1^2+y_3^2}(4c_{\beta}^2(y_1^2+y_3^2)(y_2^2-y_3^2)+s_{\beta}^2(2y_1^2-y_3^2)(2y_2^2+y_3^2))-fs_{\beta}vy_3(2y_1^2-y_2^2)(y_2^2-y_3^2)(3c_W^2\sqrt{y_1^2+y_3^2}+8s_W^2(y_1^2+y_2^2)(y_1^2+y_3^2))-2f^2s_W^2y_1y_2(y_1^2+y_2^2)(y_1^2+y_3^2)^{3/2}(y_3^2-y_2^2)] $\\
        \hline
$t \bar T Z'$ & $g^{t \bar T  Z'}_{A}=\frac{gs_\beta v(2y_1^2-y_2^2)y_3}{8f(y_1^2+y_2^2)^2(y_1^2+y_3^2)}  $\\
        \hline
$T\bar{t}h$ & $g^{T\bar{t}h}_{A}=-\frac{c_{\alpha}}{2f(y_1^2+y_2^2)^{3/2}(y_1^2+y_3^2)^{3/2}}[-(f(2y_1^4+y_1^2y_2^2-y_2^4)y_3(y_1^2+y_3^2))+vy_1\sqrt{y_1^2+y_3^2}(2y_1^4(y_2+y_3)-2y_2^2y_3(y_2+y_3)^2+y_1^2(2y_2^3+5y_2y_3^2-2y_3^3))s_{\beta}]$ \\
        \hline
		$t \bar T  H$ &   $g_A^{t \bar T H}=\frac{1}{12f^2(y_1^2+y_2^2)^{5/2}(y_2^2-y_3^2)(y_1^2+y_3^2)^{3/2}}[(v^2y_3(-y_2^2+y_3^2)(-20y_1^8+16y_2^6y_3^2-3y_2^4\sqrt{y_1^2+y_3^2}-4y_1^6(6y_2^2+5y_3^2)-6y_1^4(-2y_2^4+4y_2^2y_3^2+\sqrt{y_1^2+y_3^2})+y_1^2(16y_2^6+12y_2^4y_3^2+9y_2^2\sqrt{y_1^2+y_3^2}))\frac{c_{\alpha}s_{2\beta}}{2})+(6f^2(2y_1^2-y_2^2)(y_1^2+y_2^2)^2y_3(y_2^2-y_3^2)(y_1^2+y_3^2)-2v^2(2y_1^2-y_2^2)(y_1^2+y_2^2)y_3(y_2^2-y_3^2)(y_1^2+y_3^2)c_{\beta}^2+6fvy_1(y_1^2+y_2^2)(y_2^2-y_3^2)\sqrt{y_1^2+y_3^2}(2y_1^4(y_1+y_3)-2y_2^2y_3(y_2+y_3)^2+y_1^2(2y_2^2+5y_2y_3^2-2y_3^3))s_{\beta}-3v^2y_3(4y_1^3(2y_2^2+y_3^2)+2y_1^6(5y_2^4+11y_2^2y_3^2+2y_3^4)+y_1^2y_2^4(-6y_2^4+2y_2^2y_3^2+16y_3^4-3\sqrt{y_1^2+y_3^2})-y_2^4(-y_2^2+y_3^2)(-6y_2^2y_3^2+\sqrt{y_1^2+y_3^2})+2y_1^4(-2y_2^6+13y_2^4y_3^2+2y_3^2\sqrt{y_1^2+y_3^2}+y_2^2(7y_3^4+\sqrt{y_1^2+y_3^2})))s_{\beta}^2)s_{\alpha}]$\\
		\hline
		$T\overline{T}\gamma$    &   $g_A^{T\bar{T}\gamma}=\frac{gs_Wv^2}{6f^2(y_1^2+y_2^2)^2(y_1^2+y_3^2)(y_2^2-y_3^2)}[y_1^6(8y_2^4-y_2^2(4y_3^2+1)+5y_3^4+y_3^2)+y_1^4(8y_2^6+y_2^4(4y_3^2-2)+y_2^2y_3^2(y_3^2-3)+5(y_3^6+y_3^4))+y_1^2y_2^2(y_2^4(8y_3^2-1)+y_2^2(3y_3^2-4y_3^4)+y_3^4(5y_3^2-2))-y_1^2c_{2\beta}(y_1^4(12y_2^2y_3^2+y_2^2-3y_3^2-y_3^2)+y_1^2(2y_2^4(6y_3^2+1)+y_2^2y_3^2(9y_3^2-5)-3y_3^4(y_3^2-1))+y_2^6+y_2^4y_3^2(12y_3^2+5)-3y_2^2y_3^4(y_3^2+2))-2y_2^6y_3^2+2y_2^4y_3^4]$\\
		\hline
	$T \bar{T}Z $  &   $g_A^{T \bar{T}Z }=0 $\\
	\hline
	$T \bar{T}Z' $  &   $g_A^{T \bar{T}Z' }= -\frac{g}{2} $\\
	\hline
$Z \mu^{+}\mu^{-} $  &   $g_A^{Z \mu \mu }=\frac{g}{4c_W} $\\
	\hline
	$Z' \mu^{+}\mu^{-} $  &   $g_A^{Z' \mu \mu }= -\frac{g}{4} $\\
	\hline
	$\gamma \mu^{+}\mu^{-} $  &   $g_{A}^{\mu \mu \gamma }=0  $\\
	\hline
\end{tabular}
\end{table}

\begin{table}[H]
\caption{Effective couplings derived from interaction vertices involving a Higgs boson.}
\label{Table-i}
\begin{tabular}{|p{2.5cm}| p{13.0cm}| p{7.0cm}|c|}
\hline
\hline
  \textbf{Vertex}    &   \hspace{4.2cm}\textbf{Effective couplings} \\
\hline
\hline
		$h T \overline{T}$ & $g_{h T \overline{T}} =\frac{vy_{1}^2 s_\beta c_\alpha \left(y_{1}^4 \left(y_{2}^2+2y_{3}^2\right)+y_{1}^2 \left(8y_{2}^2 y_{3}^2-2y_{3}^4\right)-y_{2}^2 \left(y_{2}^4+3y_{2}^2 y_{3}^2-7y_{3}^4\right)\right)}{f \left(y_{1}^2+y_{2}^2\right)^{3/2}
			\left(y_{1}^2+y_{3}^2\right) \left(y_{2}^2-y_{3}^2\right)} $ \\
		\hline
		$H T \overline{T}$   &  $ g_{H T \overline{T}}= -\frac{v y_{1}^2 s_\beta s_{\alpha}\left(y_{1}^4 \left(y_{2}^2+2y_{3}^2\right)+y_{1}^2 \left(8 y_{2}^2 y_{3}^2-2 y_{3}^4\right)-y_{2}^2 \left(y_{2}^4+3y_{2}^2 y_{3}^2-7 y_{3}^4\right)\right)}{f \left(y_{1}^2+y_{2}^2\right)^{3/2}
			\left(y_{1}^2+y_{3}^2\right) \left(y_{2}^2-y_{3}^2\right)}$ \\
		\hline
		$h \mu^{+} \mu^{-} $   &   $g_{h \mu \mu }=\frac{g m_\mu}{2 m_W}$   \\
		\hline
		$H \mu^{+} \mu^{-} $   &   $ g_{H \mu \mu}= -s_{\alpha}y_{\mu}+\frac{v^2 y_\mu (-2 s_\beta c_\beta\,c_\alpha+3 s^2_{\beta}\,s_\alpha+ c^2_{\beta} s_\alpha)}{3 f^2}$  \\
		\hline
\end{tabular}
\end{table}

\newpage

\end{document}